\documentclass[lettersize,journal]{IEEEtran}
\usepackage{amsmath,amsfonts}
\usepackage{algorithmic}
\usepackage{algorithm}
\usepackage{array}

\usepackage{textcomp}
\usepackage{stfloats}
\usepackage{url}
\usepackage{verbatim}
\usepackage{cite}
\usepackage{graphicx,xcolor,tikz}
\usepackage{latexsym}
\usepackage{comment}
\usepackage{float}
\usepackage{multicol}
\usepackage{cite}
\usepackage{makecell}
\usepackage{siunitx}
\usepackage{multirow}
\usepackage{mathtools}
\usepackage{amsmath,amssymb, amsfonts}

\usepackage{tabulary}
\usepackage{booktabs}
\usepackage{mathtools}

\def\nb0{{\mathbf{0}}}
\def\nb1{{\mathbf{1}}}

\usepackage{tikz,epstopdf}

\usetikzlibrary{arrows}

\usepackage[caption=false,font=normalsize,labelfont=sf,textfont=sf]{subfig}
\usepackage{caption}
\usepackage{subfig}

\usepackage{lipsum,graphicx}
\usepackage[para]{threeparttable}
\usepackage{epsfig,graphics,graphicx,amssymb,amstext,amsmath,algorithm, algorithmic,wrapfig,multirow}
\usepackage{adjustbox}
\usepackage{pgfplots}
\usepackage{xcolor}

\usepackage[breaklinks]{hyperref}

\usetikzlibrary{chains,arrows,calc ,positioning}
\usepackage{graphicx}
\usepackage{colortbl, xcolor}
\usepackage{enumitem}
\usepackage{fancyhdr}

\usepackage{newtxtext}
\usepackage{newtxmath}
\pgfplotsset{compat=1.18}

\def\x   {{\boldsymbol{x}}}
\def\r   {{\boldsymbol{r}}}
\def\btheta   {{\boldsymbol{\theta}}}
\def\bphi   {{\boldsymbol{\phi}}}
\def\z {{\boldsymbol{z}}}
\def\u {{\boldsymbol{u}}}
\def\ISD  {{\mathsf{ISD}}}

\def\txphi {\phi^{\rm{tx}}}
\def\txtheta {\theta^{\rm{tx}}}
\def\rxphi {\phi^{\rm{rx}}}
\def\rxtheta {\theta^{\rm{rx}}}

\begin{document}

\title{A Geometry-based Stochastic Wireless Channel Model using Generative Neural Networks}

\author{Seongjoon Kang (email: sk8053@nyu.edu)
}


\IEEEpubid{0000--0000/00\$00.00~\copyright~2026 IEEE}

\maketitle

\begin{abstract}
Due to the high complexity of geometry-deterministic wireless channel modeling and the difficulty in its implementation, geometry-based stochastic channel modeling (GBSM) approaches have been used to evaluate system performance of wireless communications. 
This paper introduces a new method to model a GBSM by training a generative neural network using images formed by channel parameters.
Toward this end, we process the data of channel parameters in the form of images and train the generative neural networks where the convolutional layers are mainly employed to capture correlation among multipath components. 
\textcolor{black}{
Through a case study, we demonstrate that the use of channel images facilitates the training of the generative model and ensures that the model learns the correlations among multipath components. We show that the outputs of the generative model faithfully represent the joint distributions of the original data, and that the trained model reliably interpolates across held-out conditions not used during training, demonstrating its practical value as a data-efficient alternative to directly resampling the ray-tracing database.}
Furthermore, to corroborate applicability of the trained model, we run simple system-level simulations and show the results obtained from the trained model closely match those from the ray-tracing data.
 Therefore, the proposed model is expected to ease the burden of GBSM implementations with general wireless conditions and capture the statistical joint distributions of the original channel data. 
\end{abstract}


\section{INTRODUCTION}

The goal of communication channel modeling is to facilitate link- and system-level simulations in which the performance of the wireless system is assessed. The $3$rd Generation Partnership Project ($3$GPP) proposes different types of stochastic channel model depending on wireless systems such as aerial and satellite communications for benchmarking \cite{3GPP38901, 3GPP36777, 3GPP38811}.

The statistical channel models proposed by $3$GPP aim at general scenarios such as urban micro- and rural regions and do not take into account geometry-specific scenarios. 
To consider an underlying geometry, deterministic models, such as using ray-tracing simulation, have been proposed on the basis of fundamental electromagnetic wave propagation theory. In particular, Wireless Insite~\cite{Remcom} has been known to be fairly accurate compared to real-world measurements, as shown in~\cite{de2023ray, remcommpresnt2017modeling}.
However, due to the high computational and implementation cost of deterministic models, geometry-based stochastic channel modeling (GBSM) is proposed \cite{yin2016propagation}, which models channel parameters through statistical distributions of predefined local scatterers in certain environments.

GBSMs simplify the ray-tracing rules without specific information of local scatterers. Thus, GBSMs can offer more flexibility and convenience to implement and perform simulations in various scenarios due to their stochastic characteristics.
In \cite{akki1986statistical}, the first GBSM was proposed by developing probabilistic distributions of channel parameters, and in \cite{patzold2008modeling, cheng2009adaptive, zajic2008space}, two-ring scattering models are employed using wave propagation rules in random scatterers. In \cite{wallace2001statistical, li2002impact}, the clustered models are proposed for parameterized channel modeling. 
In addition, several efforts have been made to model GBSMs using real measurement data \cite{zhou2019geometry,karedal2009geometry, roivainen2016geometry}.

However, prior efforts for parameterization of GBSMs still require the high complexities of implementation and do not generalize the modeling under any wireless condition. Thus, it is essential to explore ways to reduce the implementation burden of GBSMs and ensure almost the same modeling accuracy as ray-tracing modeling. As proposed in \cite{yang2018geometry}, data obtained from the ray-tracing simulation can be used to fit the parameters of the GBSMs and guarantee good modeling accuracy. 
\IEEEpubidadjcol
 In efforts to build data-driven GBSMs, generative neural networks are used \cite{yang2019generative, orekondy2022mimo, xia2022generative, hu2022multi, xiao2022channelgan, seyedsalehi2019propagation}. For example, the authors of \cite{xia2022generative} propose the conditional variational autoencoder (CVAE) to capture the channel parameter statistics. This work requires two-step procedures---link state classifier and VAE---to stochastically generate channel parameters, which still require higher implementation complexity. Furthermore, this effort is based on only fully connected (FC) neural networks, and thus does not consider the correlation between multipath components, i.e., \emph{the joint distribution} of all the multipath components. \textcolor{black}{In particular, training CVAE is challenging due to the posterior collapse problem \cite{mccarthy2020addressing}, where the latent variable is ignored and the model degenerates into a standard autoencoder, thereby losing the ability to generate diverse channel realization}.
 \textcolor{black}{Following the same implementation procedures and relying solely on FC neural networks, \cite{hu2022multi} proposes a cluster-based generative model built upon Wasserstein generative adversarial networks with gradient penalty (WGAN-GP). However, generating channel parameters at the cluster level leads to significant limitations in simulation accuracy. This is because wideband channels are characterized by finer time resolution, which necessitates modeling a larger number of resolvable multipath components than cluster-level generation can reliably capture.}

Therefore, it is desirable to simplify and generalize the implementation of GBSMs using a generative model to capture not only the statistics of the parameters of multipath channels, such as pathloss and propagation delay, but also \emph{correlations} between them. 
Toward this end, the contributions of this work are as follows. 
\textcolor{black}{
\begin{itemize}
    \item We propose a novel data-to-image mapping that converts wireless channel parameters into \emph{channel images}, enabling generative models to be trained with two-dimensional convolutional neural networks (2D CNNs). By exploiting the texture bias of CNNs, the proposed channel images allow the model to capture \emph{the joint distribution of multipath components}, which prior FC-based generative channel models~\cite{xia2022generative, hu2022multi} fail to reproduce.
    \item Through a case study based on ray-tracing data of a dense urban area, we validate the statistical fidelity of the proposed model in terms of the EM distance, the distributions of individual channel parameters, and the correlations among multipath components. In particular, comparisons with the FC-based baselines~\cite{xia2022generative, hu2022multi} demonstrate that the proposed model reproduces the correlation structure of the original data, and further exhibits statistical interpolation capability under unseen 2D distance conditions.
    \item We corroborate the applicability of the proposed model to link- and system-level evaluations by comparing drop-based simulation results with those obtained from the ray-tracing channels. The close agreement between the SNR distributions of the two channel models confirms that the proposed model can replace costly ray-tracing simulations for site-specific performance analysis.
\end{itemize}}

This paper is organized as follows. Section~\ref{sec:2_wireless_chan} describes the general wireless channel modeling and the consequential required channel parameters, which motivates this work. Section~\ref{sec:3_chan_data_process} sets forth ways to process channel data, and Section~\ref{sec:4_design_cond_chan} presents a conditional generative model for GBSMs. To verify our proposed modeling method, Sections~\ref{sec:5_case_study} and~\ref{sec:6_model_train_recover} introduce a case study and explain the training of the model. Then, in Section~\ref{sec:7_model_eval}, the performance of the trained model is assessed considering various evaluation metrics. Lastly, Section~\ref{sec:8_application_model} discusses the application of the proposed model in the aspect of the performance of system-level simulation, and Section~\ref{sec:9_conclusion} concludes this work. 
\section{GENERAL WIRELESS CHANNEL MODELING}
\label{sec:2_wireless_chan}
In current 5G and future 6G communications, multiple-in multiple-out (MIMO) systems between transmitter (TX) and receiver (RX) have become common; thus, the wireless channels of MIMO systems are of paramount importance. 
In most cases, MIMO channels are modeled with their spatial signatures \cite{tse2005fundamentals,heath2018foundations}. Denoting the number of antennas of TX and RX as $n_{\rm{tx}}$ and $n_{\rm{rx}}$, MIMO channel matrix $\boldsymbol{H}\in \mathbb{C}^{n_{\rm{rx}} \times n_{\rm{tx}}}$ between the TX and RX is described as follows:
\begin{align}
    \boldsymbol{H}(t) &= \sum_{k=1}^N \sqrt{p_k^{-1}F_{\rm{tx}}(\txtheta_k, \txphi_k)F_{\rm{rx}}(\rxtheta_k, \rxphi_k)} e^{\psi_k i} \nonumber \\ 
    &~\boldsymbol{e}_{\rm{rx}}(\rxtheta_k, \rxphi_k) \boldsymbol{e}_{\rm{tx}}^{\rm{H}}(\txtheta_k, \txphi_k) \delta (t-\tau_k) 
    \label {eq:mimo_chan}
\end{align}
where $t$ is the measured time after transmission,  $N$ is the total number of resolvable multipaths, for $k$th path, $p_k$ is pathloss, $F_{\rm{tx}} (\cdot, \cdot)$ and $F_{\rm{rx}}(\cdot, \cdot)$ are antenna element gains for TX and RX, $\psi_k$ is arrival phase, $\delta$ is Dirac delta function, $\tau_k$ is propagation delay,  $\txphi_k$ and $\txtheta_k$ are azimuth and zenith angles of departure (AOD and ZOD), $\rxphi_k$ and $\rxtheta_k$ are azimuth and zenith angles of arrival (AOA and ZOA), and $\boldsymbol{e}_{\rm{tx}} (\cdot, \cdot) \in \mathbb{C}^{n_{\rm{tx}} \times 1}$ and $\boldsymbol{e}_{\rm{rx}}(\cdot, \cdot) \in \mathbb{C}^{n_{\rm{rx}} \times 1}$ are the spatial signatures from antenna arrays of TX and RX. 

By taking the Fourier transform of Eq.~\ref{eq:mimo_chan}, the channel frequency response $\boldsymbol{H}(f)$ for a MIMO system can be derived, which is essential to model channels of orthogonal frequency division multiplexing (OFDM) over all available carrier frequencies. When ignoring the mutual coupling effect between the antenna elements, Eq.~\ref{eq:mimo_chan} is sufficient to describe all the channel coefficients between two antenna arrays of TX and RX.

However, to stochastically build MIMO channels we should stochastically obtain all the necessary parameters in Eq.~\ref{eq:mimo_chan}. Thus, a stochastic model is required to generate all the parameters in Eq.~\ref{eq:mimo_chan} such as pathloss and arrival and departure angles. To this end, we employ generative neural networks to probabilistically obtain all the essential parameters in Eq.~\ref{eq:mimo_chan} under a certain condition.  
\section{CHANNEL DATA PROCESSING}
\label{sec:3_chan_data_process}
In this section, we discuss how to process the raw data of channel parameters before training a generative model. 
\label{sec:data_processing}
\subsection{Channel Parameters From Ray-Tracing Simulation}
Since measuring realistic wireless channels in any region is challenging, we rely on the simulation and obtain the data from it.
The parameters of the wireless channel for a specific region can be obtained from commercial \emph{ray tracer}, Wireless Insite\cite{remcomm}. For each wireless link, we consider all possible propagation path parameters, which are the resultant outputs from ray-tracing simulator. With those parameters, we can model any wireless channel.

Specifically, for each link consisting of $N$ paths between a TX and an RX, the channel parameter matrix $\boldsymbol{D} \in \boldsymbol{R}^{8 \times N}$ obtained from the ray-tracing simulator is represented below.
\begin{equation} \boldsymbol{D} =
\left (
\begin{array}{cccc}
{p}_1 & {p}_2 & \ldots & {p}_N\\
{\tau}_1 & {\tau}_2 & \ldots & {\tau}_N\\
{\txphi_1} & {\txphi_2} & \ldots & {\txphi_N}\\
{\txtheta_1} & {\txtheta_2} & \ldots & {\txtheta_N}\\
{\rxphi_1} & {\rxphi_2} &  \ldots & {\rxphi_N}\\
{\rxtheta_1} & {\rxtheta_2} & \dots & {\rxtheta_N}\\
{\psi_1} & {\psi_2} &  \dots & {\psi_N}\\
{l}_1 & {l}_1 & \ldots & {l}_1 \\
\end{array}
\right )
\label{eq:data_mat}
\end{equation}
where $l_1$ is the link state of the first arrival path. Here link state indicates the state of communication link between TX and RX, which is normally classified as three states: line-of-sight (LOS), non-line-of-sight (NLOS), and outage. Note that we only need the link state of the first arrival path since the other paths are NLOS paths. The single-matrix representation for all channel parameters on one link is useful in building channel images. 
The LOS path components, which constitute the first column of $\boldsymbol{D}$, are determined mathematically as follows:
{\allowdisplaybreaks
\begin{subequations}
\begin{align}
    &p_1 = 20\log_{10}({\rm dist}_{\rm 3d})+20\log_{10}(f_c)-147.55 \label{eq:los_pl}\\
    &\tau_1 = \frac{{\rm dist}_{3\rm d}}{c} \label{eq:dly}\\
    &\txphi_1 = \arctan{\frac{y_{\rm tx}-y_{\rm rx}}{x_{\rm tx} - x_{\rm rx}}} \label{eq:aod}\\
    &\rxphi_1  =  \txphi_1 - 180 \label{eq:aoa}\\
    &\txtheta_1 = \arctan{\frac{{\rm dist}_{\rm 2d}}{z_{\rm rx}-z_{\rm tx}}} \label{eq:zod}\\
    &\rxtheta_1 = 180 - \txtheta_1 \label{eq:zoa}\\
    &\psi_1 = (f_c \cdot \tau_1)~~\rm{mod}~~360
    \label{eq:los_ps}
\end{align}
\end{subequations}}
where $\rm {dist}_{3d}$ is $3$D distance between TX and RX, $f_c$ is the carrier frequency, $c$ is the speed of light, ($x_{\rm tx}$, $y_{\rm tx}$) and ($x_{\rm rx}$, $y_{\rm rx}$) are the $2$D coordinates of the TX and the RX. 

\subsection{Data Normalization}
In order to train a generative neural network, the raw data obtained from the ray-tracing simulation are normalized by employing \emph{Min-Max} scaling to make the range of all elements of the data matrix 
 $\boldsymbol{D}^i \in \boldsymbol{R}$ fall between -1 and 1. This data pre-processing is commonly employed to reduce variance.

Suppose that the $M$ TX-RX links are simulated in ray-tracing simulation. Since each link has a different number of paths, the corresponding data matrices may have different data shapes. 
To ensure that the size of the data matrix is consistent across all links, 
we introduce \emph{virtual paths} when the number of paths is smaller than the maximum value of 25.
 Except for the pathloss feature, the virtual values of the feature data $\boldsymbol{d}_i$ are sampled from the uniform distribution $U(\rm{\min}~\boldsymbol{d}_i,\rm{\max}~\boldsymbol{d}_i)$.
 \textcolor{black}{
 For the pathloss feature, we deliberately select virtual values exceeding $\SI{180} {\decibel}$---e.g., sampled from $U(180, 200)$---as these values serve as the marker distinguishing virtual paths from real ones.}


Before Min-Max scaling for the raw data, we take into account the following data processing technique for some features.
\begin{itemize}
    \item {Pathloss}: Since we assume that the TX and RX locations are known, we normalize the pathloss values per path by the free space pathloss (FSPL) using Eq.~\ref{eq:los_pl}
    \begin{equation*}
        \Tilde{\boldsymbol{p}} = \boldsymbol{p} - FSPL(\rm {dist}_{3d})
    \end{equation*}
    \item {Delay}: Similarly, the delay values are normalized by the LOS delay using Eq.~\ref{eq:dly} 
    \begin{equation*}
        \Tilde{\boldsymbol{\tau}} = \boldsymbol{\tau} - \frac{\rm {dist}_{3d}}{c}
    \end{equation*}
    Considering that delay values are relatively small compared to other feature data, we multiply them by a sufficiently large scalar value---$10^7$.
    \item{Link state}:  
     The link state values of the first arrival path are sampled from uniform distributions, which are $U (1-\epsilon, 1)$ for LOS and $U(-1, -1+\epsilon)$ for NLOS, respectively, for a small value $\epsilon$ ($0<\epsilon\ll1$).
\end{itemize}

 We do not consider angles relative to the LOS directions because this processing might not be invertible under certain conditions.
 For example, when the conditional variables are the 2D distance between a TX and RX and the height of an RX, the AOD and AOA are not recoverable from angles relative to the LOS direction. 

Denoting the total data tensor as $\mathcal{\boldsymbol{D}}^{\rm {total}} = \{\boldsymbol{D}^1, \boldsymbol{D}^2, \ldots, \boldsymbol{D}^M \}\in \boldsymbol{R}^{M \times 8 \times 25}$,
the Min-Max scaling for each feature data is given as follows:
\begin{align*}
    \hat{\boldsymbol{d}_i} &= \frac{(2 \times  \boldsymbol{d}_i - \max{\boldsymbol{d}_i} - \min{\boldsymbol{d}_i})}{\max{\boldsymbol{d}_i} - \min{\boldsymbol{d}_i}}\\
    \boldsymbol{d}_i &= \mathcal{\boldsymbol{D}}^{\rm {total}}_{:,i,:}
\end{align*}
Note that all the pre-processing for raw data must be invertible, since we need the reverse process to reconstruct data from the output of a generative model. 

\begin{figure}[!b]
\centering
\subfloat[][$25$ paths]
{\includegraphics[width =0.45\columnwidth]{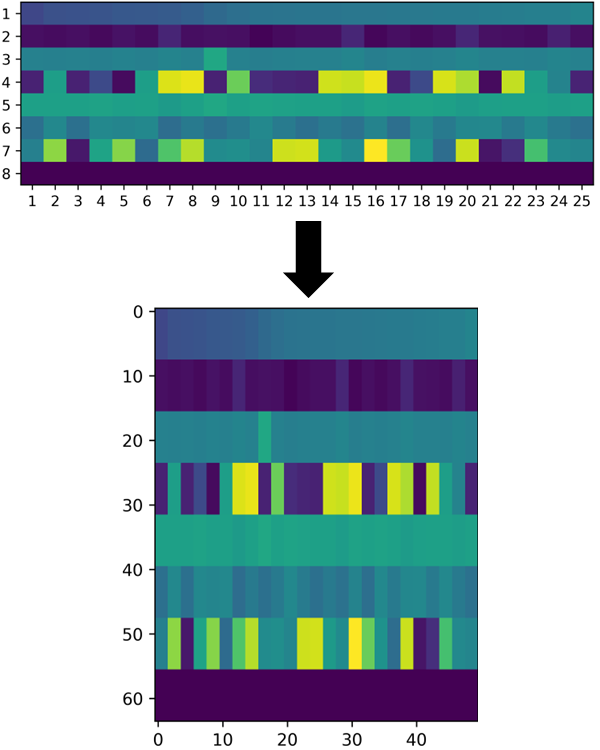} \label{fig:25paths}}
\subfloat[][$13$ paths]
{\includegraphics[width =0.45\columnwidth]{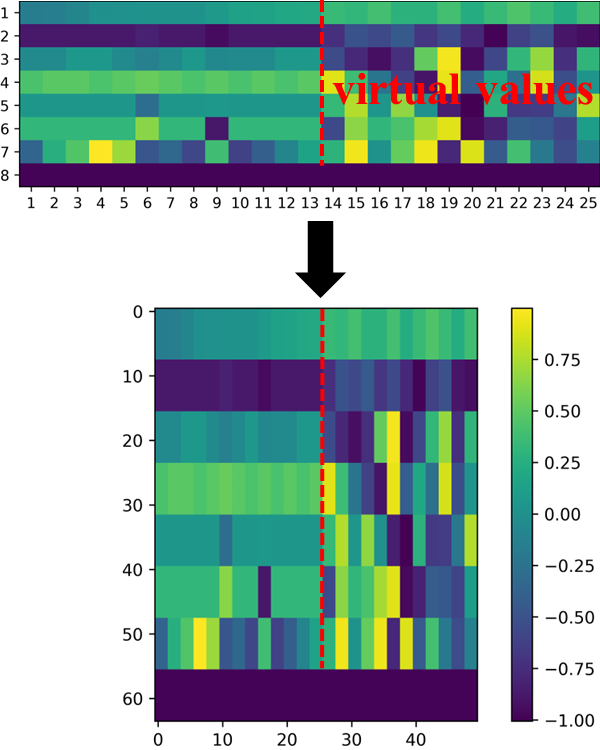} \label{fig:13paths}}
\caption{Images for different number of multipaths are generated by duplicating each element of the pre-processed channel matrix horizontally and vertically.}
\label{fig:images_from_two}
\end{figure}

\subsection{Data-To-Image Mapping}
\label{sec:data_to_image_mapping}
Following data pre-processing, we generate channel images for each wireless link between TX and RX. The formation of images is instrumental in capturing \emph{correlations} between multipath components by employing $2$D CNN. \textcolor{black}{Unlike natural images, wireless channel data inherently lack macro-level geometric shapes or continuous contours. When a two-dimensional  convolutional kernel is applied directly to the raw $8 \times 25$ channel matrix, heterogeneous channel parameters are inevitably mixed within a single receptive field, preventing the network from learning discriminative channel parameters. Thus, we propose duplication-based preprocessing by enlarging each element into a homogeneous block, ensuring that the initial convolutional operations are confined within identical physical parameters, as illustrated in Fig. ~\ref{fig:images_from_two}. This duplication-based transformation introduces spatially repetitive patterns consistent with the textural features where CNNs are inherently biased to learn \cite{geirhos2018imagenet}. By exploiting this texture bias, the network can more effectively capture discriminative channel features, improving the overall performance of the generative model.}

Specifically, we duplicate each element of the data matrix $\boldsymbol{D}^i \in \mathbb{R}^{8 \times 25}$ by $2$ times horizontally, $8$ times vertically,  so that the size of the resulting matrix becomes $\Tilde{\boldsymbol{D}}^i \in \mathbb{R}^{64 \times 50}$. 
\begin{equation*}
    \boldsymbol{D}^i \in \mathbb{R}^{8 \times 25} \rightarrow \Tilde{\boldsymbol{D}}^i_{\rm image} \in \mathbb{R}^{64 \times 50}
\end{equation*}
Fig.~\ref{fig:images_from_two} illustrates the process of generating images from the original data matrices for two different multipath scenarios.
As shown in Fig.~\ref{fig:images_from_two}(b), when the number of multipaths is $13$ ($<25$), the virtual values are filled to make the size of the data matrix consistent. 
 \textcolor{black}{
Note that the virtual paths are introduced solely for training purposes, to equalize the dimension of the data matrices, and are discarded after data generation based on the pathloss threshold of $\SI{180}{\decibel}$. The virtual values are randomly sampled, rather than set to constants, because constant padding would create large uniform regions without textural variation, preventing the network from exploiting the texture bias of CNNs.}
In addition, we duplicate the image channel by $3$ times, and thus the final image size is $\Tilde{\boldsymbol{D}}^i_{\rm image} \in \mathbb{R}^{3 \times 64 \times 50}$


\section{DESIGN OF CONDITIONAL GENERATIVE MODEL}
\label{sec:4_design_cond_chan}
The objective of the proposed GBSM is to capture channel statistics for a specific region that corresponds to certain conditions, such as RX heights, distances between TX and RX, and carrier frequencies. Therefore, we need to design a generative model that can conditionally output the channel parameters. 
There exist various types of generative models such as VAE and diffusion model, but in this work we employ Wasserstein generative adversarial networks (WGAN) \cite{arjovsky2017wasserstein}.

\subsection{Overview Of WGAN-GP}
Generative adversarial networks (GANs) \cite{goodfellow2014generative} have been a popular generative model for many years, but face several challenges, such as training instability and mode collapse. For this reason, WGANs were introduced to resolve the issues of traditional GANs \cite{arjovsky2017wasserstein}. 
Unlike the KL divergence, which suffers from vanishing gradients when the two distributions have disjoint supports, WGAN adopts the Wasserstein-1 distance, also known as the Earth Mover (EM) distance, to ensure smoother and more informative gradients throughout training. The EM distance is made tractable via the Kantorovich-Rubinstein duality \cite{villani2009optimal}, which yields
\begin{align}
W\left(\mathbb{P}_r, \mathbb{P}_\btheta\right)=\sup _{\|f\|_L \leq 1} \mathbb{E}_{\x \sim \mathbb{P}_r}[f(\x)]-\mathbb{E}_{\x \sim \mathbb{P}_\btheta}[f(\x)]
\end{align}
where $\btheta$ denotes the parameters of generator, $\mathbb{P}_r$ and $\mathbb{P}_\theta$ are the distributions of real data and outputs from generator, and the supremum is over all the 1-Lipschitz functions $f$, $\|f\|_L \leq 1$. Therefore, the loss of critic is defined as 
\begin{align}
   L_{\bphi}^{C} = - \left(\mathbb{E}_{\x \sim \mathbb{P}_r}\left[f_\bphi(\x)\right]-\mathbb{E}_{\z \sim p(\z)}\left[f_\bphi \left(g_\btheta(\z)\right) \right]\right)
   \label{eq:loss_disc}
\end{align}
where $\bphi$ indicates the parameters of the critic that we should train, $f_{\bphi}$ is the critic function having parameter $\bphi$,  and $\z$ is a latent variable following standard normal distribution. Theorem $3$ in \cite{arjovsky2017wasserstein} shows that the optimal parameter $\btheta$ for the generator satisfying Eq.~\ref{eq:loss_disc} can be found by 
\begin{align}
    \nabla_\btheta W\left(\mathbb{P}_r, \mathbb{P}_\btheta\right)=-\mathbb{E}_{\z \sim p(\z)}\left[\nabla_\btheta f_{\bphi}\left(g_\btheta(\z)\right)\right]
    \label{eq:gradient_theta}
\end{align}
where $g_{\btheta}$ is generator function having parameter $\btheta$.
Thus, the loss of generator is simply defined as
\begin{align}
    L_{\btheta}^{G} = -\mathbb{E}_{\z \sim p(\z)}\left[ f_{\bphi}\left(g_\btheta(\z)\right)\right]
    \label{eq:gen_loss}
\end{align}

The authors in \cite{gulrajani2017improved} propose one more condition to find the optimal critic $f_{\bphi}$ using the fact that the gradient norm of $f_{\bphi}$ at the interpolated point between the generated and the real data samples is equal to one. 
Accordingly, the new critic loss with an appended gradient penalty term is defined as follows
\begin{align}
    L_{\bphi}^{C} = &-\left(\mathbb{E}_{\x  \sim \mathbb{P}_r}\left[f_\bphi(\x)\right] -\mathbb{E}_{\z \sim p(\z)}\left[f_\bphi \left(g_\btheta(\z)\right) \right] \right) \nonumber \\
    & + \lambda \underset{\hat{\x} \sim \mathbb{P}_{\hat{\x}}}{\mathbb{E}}\left[\left(\left\|\nabla_{\hat{\x}} f_{\bphi}(\hat{\x})\right\|_2-1\right)^2\right]
    \label{eq:critic_loss}
\end{align}
where $\hat{\x} = t  \x + (1-t) g_{\btheta}(\z) $  for $0\leq t \leq 1$. We will reduce the loss values given by Eq.~\ref{eq:gen_loss} and Eq.~\ref{eq:critic_loss} to find the optimal parameters $\btheta$ and $\bphi$ of the generator and critic in WGAN. 
\begin{figure}[!t]
\centering
\includegraphics[width =0.99\columnwidth]{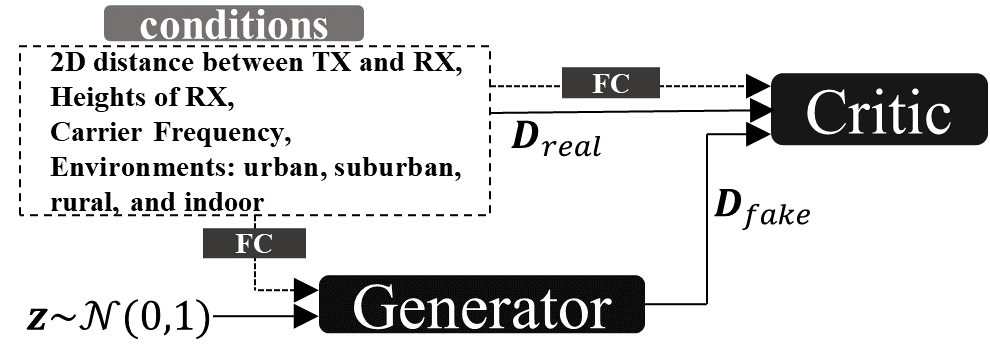} 
\caption{WGAN-GP with general wireless conditions }
\label{fig:w_gan}
\end{figure}

\begin{table}[!b]
\begin{center}
\caption{Architectures of Generator and Critic}
\label{tab:archtecture_gen_critic}
\begin{tabular}{|lllll|}
\multicolumn{5}{c}{ Generator $G(\z)$} \\
\hline & Kernel size & Resample & Output shape & Channels\\
\hline$\z $ & - & - & 25 & - \\
Conv & {$[3 \times 3] $} & $\mathrm{Up}$ & $3 \times 3$ & $512$\\
Conv & {$[6 \times 4]$} & $\mathrm{Up}$ & $8 \times 6$ & $256$\\
Conv & {$[4 \times 4]$} & $\mathrm{Up}$ & $16 \times 12$ & $128$ \\
Conv & {$[4 \times 5]$} & $\mathrm{Up}$ & $ 32 \times 25$ & $64$\\
Conv  & $[4 \times 4]$ & $\mathrm{Up}$ & $64 \times 50$ & $3$\\

\hline \multicolumn{5}{c}{ Critic $C(\x, G(\z))$} \\
\hline
Conv & {$[3 \times 3] $} & $\mathrm{Down}$ & $32 \times 25$ & $64$\\
Conv & {$[4 \times 4]$} & $\mathrm{Down}$ & $16 \times 12$ & $128$\\
Conv & {$[4 \times 4]$} & $\mathrm{Down}$ & $8 \times 6$ & $256$ \\
Conv & {$[4 \times 4]$} & $\mathrm{Down}$ & $ 4 \times 3$ & $512$\\
Conv & $[5 \times 4]$ & $\mathrm{Down}$ & $1 \times 1$ & $1$\\
\hline
\end{tabular}
\end{center}
\end{table}
\subsection{Structure of Conditional WGAN-GP}
To conditionally generate channel parameters, we concatenate conditional information to the inputs of the generator and critic using fully connected (FC) networks. 
 Fig.~\ref{fig:w_gan} describes the structure of WGAN-GP with general wireless conditions, such as 2D distances, heights, carrier frequencies, and network environments, which can be embedded in both the generator and the critic, respectively, through FCs. As we mentioned earlier, we can increase the conditional variables with the functions of the original conditions.

Table~\ref{tab:archtecture_gen_critic} summarizes the structure of the convolutional layers consisting of generator and critic, respectively. 
The latent variable $\boldsymbol{z}$ and the embedded conditional variables serve as the input for the generator, whereas the critic's input consists of the real image, the generated image, and embedded conditional variables.
As an activation function between the convolutional layers, the {LeakyReLU} function is used after some normalization process.
All detailed implementations are available at \cite{chanmod-github}.

 \section{Case Study and Ray-Tracing Simulation}
 \label{sec:5_case_study}
 In this section, we discuss the case study on GSBM with conditional constraints and explain the corresponding ray-tracing simulation to obtain the wireless channel parameters in the given region. Note that the main assumption throughout this work is that the TX and RX locations are known. 
 
 \begin{figure}[!b]
\centering
\includegraphics[width =0.99\columnwidth]{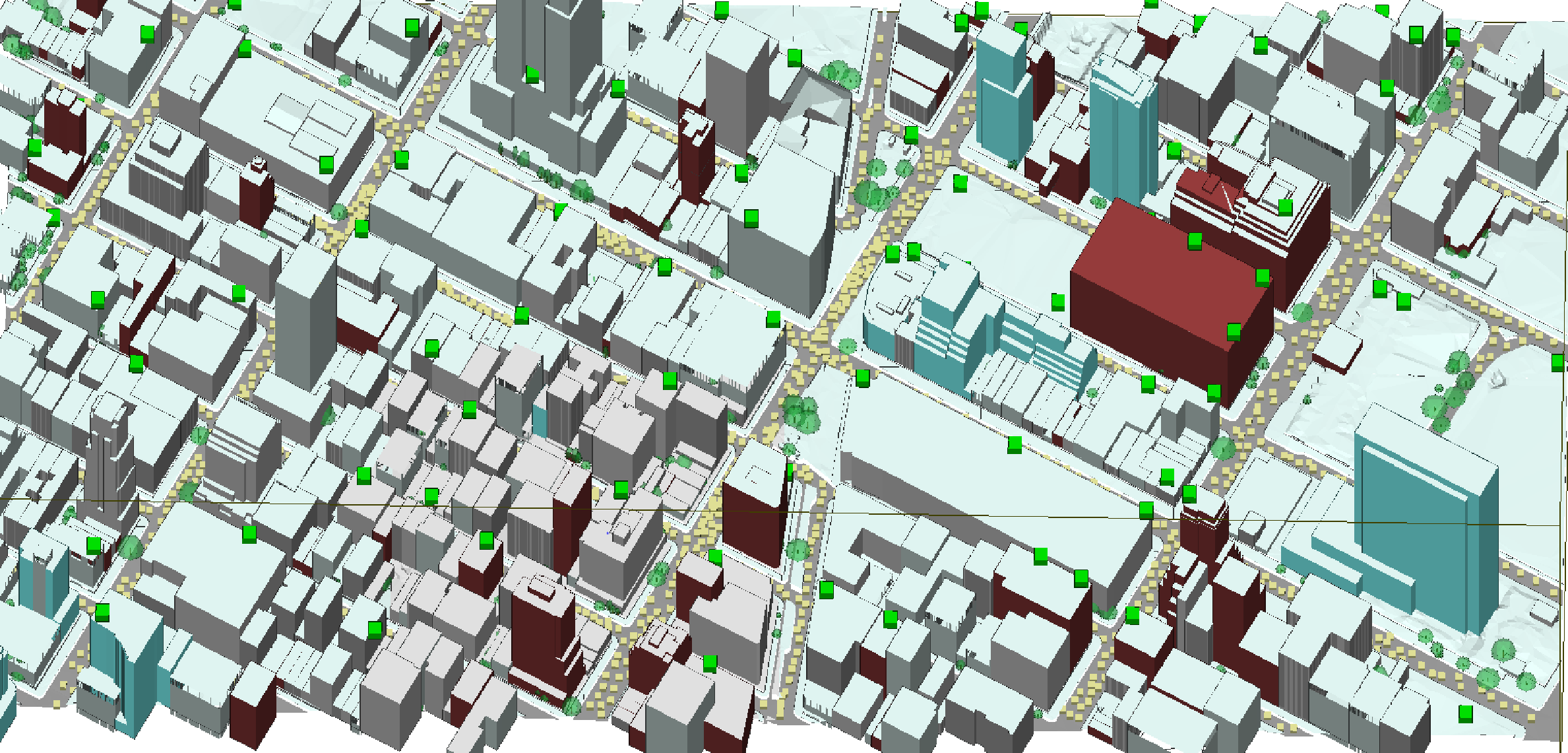} 
\caption{Herald Square in New York City for ray-tracing simulation. ($\SI{1120}{m} \times \SI{510}{m}$). The green rectangles represent TXs, and the yellow rectangles are RXs. }
\label{fig:boston}
\end{figure}

\subsection{Case Study: Varying RX Heights and 2D Distances}
As a case study, we want to capture the statistical distribution of local scatterers at different 2D distances between TXs and RXs. In addition, we take into account different RX heights with the carrier frequency fixed. Thus, the learning process of the statistical distribution of the data matrix $\boldsymbol{D}$ can be described as follows:

\begin{align}
 P(\boldsymbol{D}|\text{dist}_{\rm 2d}, \text{h} ) =  P (T \cdot G( \text{dist}_{\rm 2d}, \text{h}))
 \label{eq:cond_h}
\end{align}
where $G$ is conditional generative model---WGAN-GP in our work, and $T$ is an operator to perform image-to-data mapping process, i.e., the process from outputs of the model to the original data matrix $\boldsymbol{D}$. The correlations between the elements of the matrix $\boldsymbol{D}$
are captured through the data-to-image mapping process and convolutional operations.

We can add more conditional variables as functions of 2D distance ${\rm dist}_{\rm 2d}$ and height h, e.g., 3D distances and zenith angles.   
In this study, the five \emph{discrete} RX heights, $\SI{1.6}{m}$, $\SI{30}{m}$, $\SI{60}{m}$, $\SI{90}{m}$, and $\SI{120}{m}$, are considered as the conditions of  terrestrial and aerial channels simultaneously. Note that $2$D distance values obtained from all pairs of TX and RX are \emph{continuous} in the range of the network area.

\subsection{Ray-Tracing Simulation} 
To obtain channel parameters for the case study, we run a ray-tracing simulation on Herald Square in New York City, which can be considered a typical dense urban area.
The carrier frequency of $\SI{12}{GHz}$ is chosen in the upper midband, known for its potential as a promising spectrum for future networks \cite{fcc2023preliminary, fcc20236Gworkinggroup, kang2024cellular}. All electrical properties, including permeability and conductivity, at $\SI{12} {GHz}$ are specified for each material following the standard values specified in the International Telecommunication Union (ITU) \cite{itu527_2021,  itu2040_2021}.

\begin{table}[!b]
\begin{center}
\caption{Hyper-parameters to train WGAN-GP}
\label{tab:gan_hyperparams}
\begin{tabular}{|c|c|c|}
\cline { 2 - 3 } \multicolumn{1}{c|}{} & \begin{tabular}{c} 
 \textbf{Generator} \\
\end{tabular} & \begin{tabular}{c} 
 \textbf{Critic} \\
\end{tabular} \\
\hline Input image size & \multicolumn{2}{|c|}{[64, 50 ,3]} \\
\hline Learning rate & \multicolumn{2}{|c|} {$ 10^{-4}$} \\
\hline Optimizer & \multicolumn{2}{|c|}{ Adam ($\beta_1 = 0.5$, $\beta_2 = 0.9$)} \\
\hline Epochs & \multicolumn{2}{|c|}{20} \\
\hline Batch size & \multicolumn{2}{|c|}{256} \\
\hline
\end{tabular}
\end{center}
\end{table}

\begin{figure}[!b]
\centering
\includegraphics[width =0.99\columnwidth]{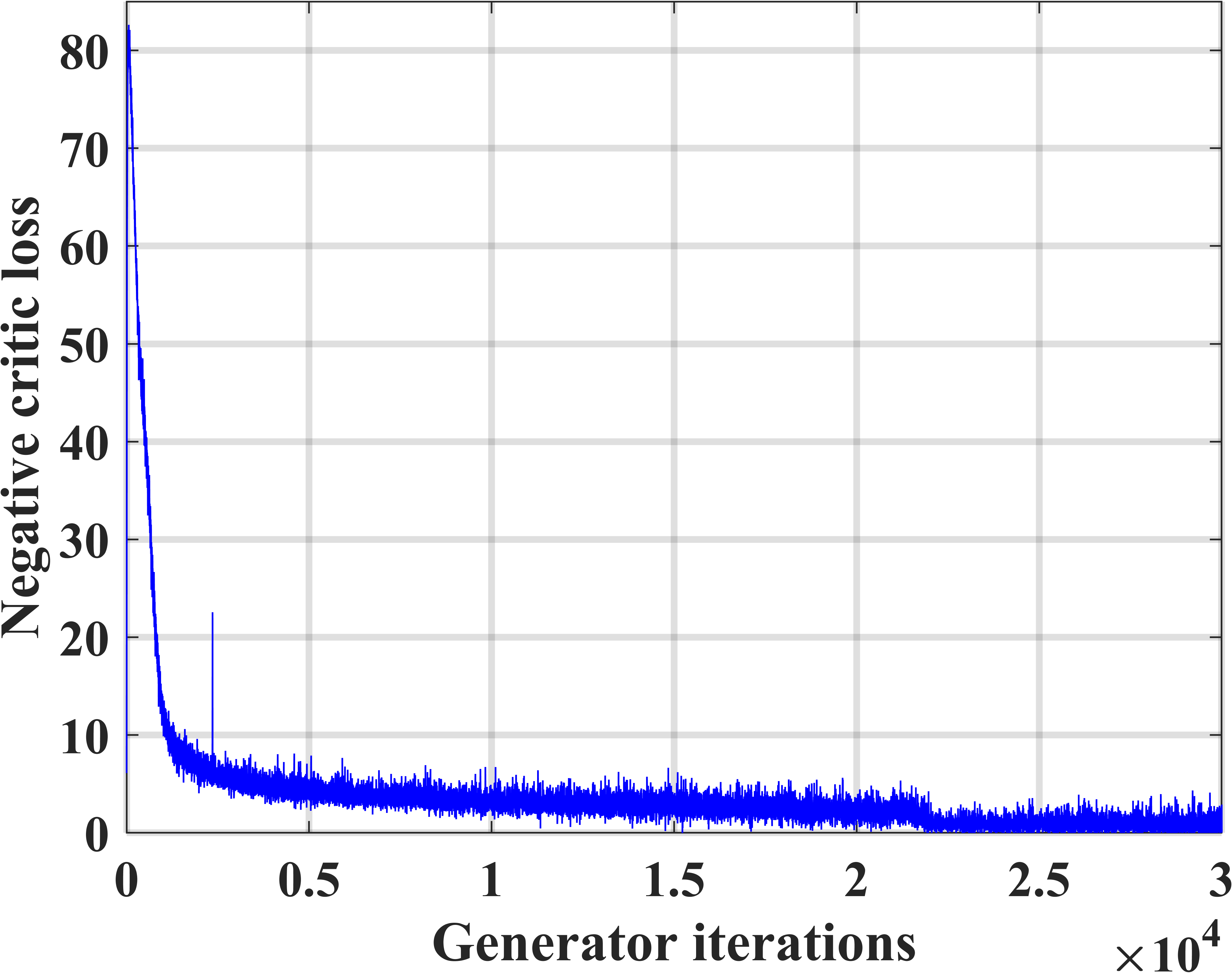} 
\caption{The negative critic loss of the proposed model}
\label{fig:critic_loss}
\end{figure}

Fig.~\ref{fig:boston} shows the area, Herald Square in New York City, for the ray-tracing simulation where TXs are placed on top of buildings and RXs are dropped on the outdoor street randomly. 
In Fig.~\ref{fig:boston}, different colors indicate different materials. For example, gray is concrete, light blue is glass, and brown is brick. To capture transmitted rays in all directions, TXs and RXs are equipped with single \emph{isotropic} antennas. 
In this simulation, we deploy $79$ TXs on top of buildings and $1526$ outdoor RXs for five different heights: 1.6, 30, 60, 90, $\SI{120}{m}$. Thus, in total $79\times5\times1526 = 602,770$ links are obtained, which can be used to model integrated aerial-terrestrial wireless channels. 
\begin{figure*}[!t]
\centering
\subfloat[][]
{\includegraphics[width =0.42\columnwidth]{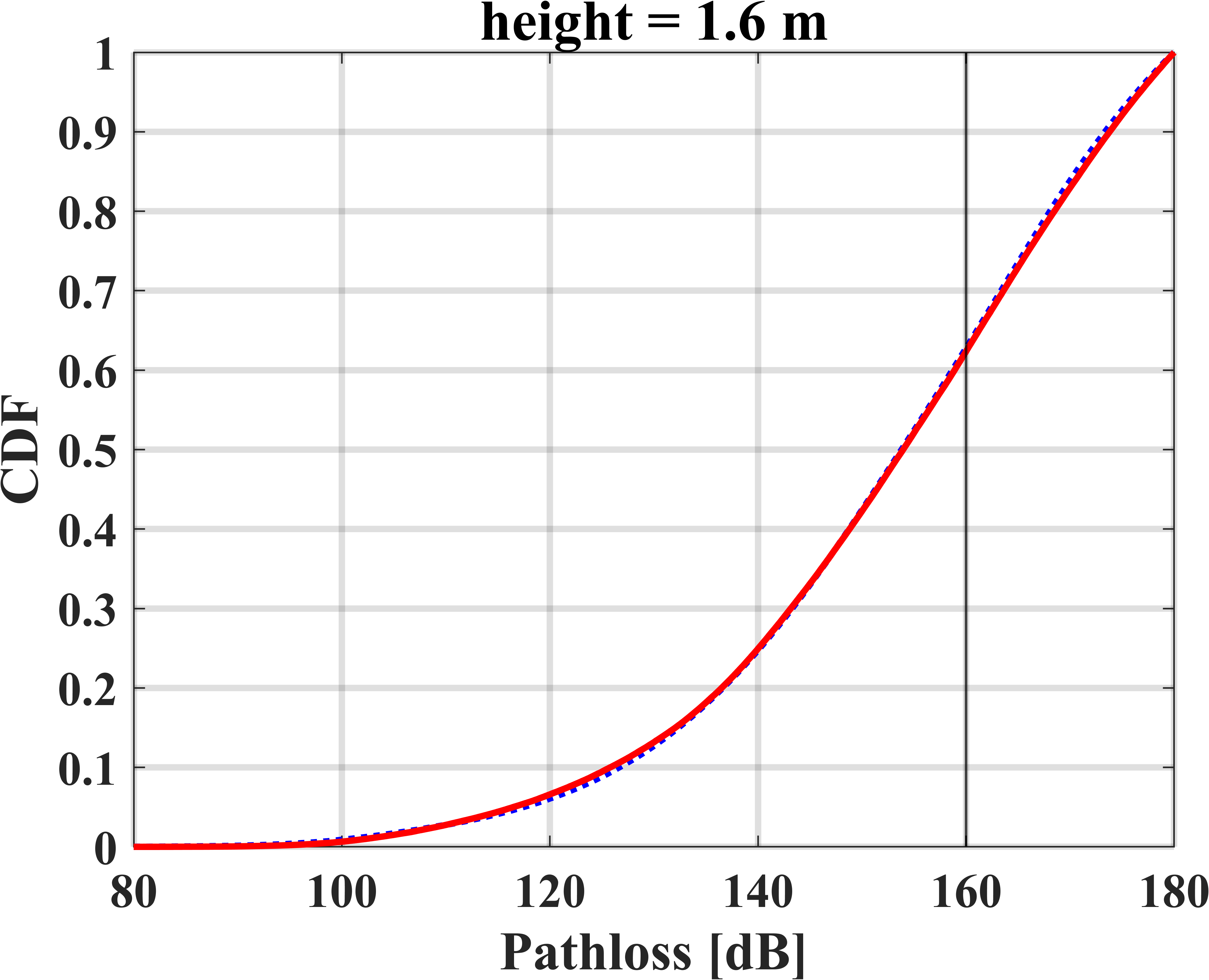} \label{fig:pl_1_6}}
\subfloat[][]
{\includegraphics[width =0.38\columnwidth]{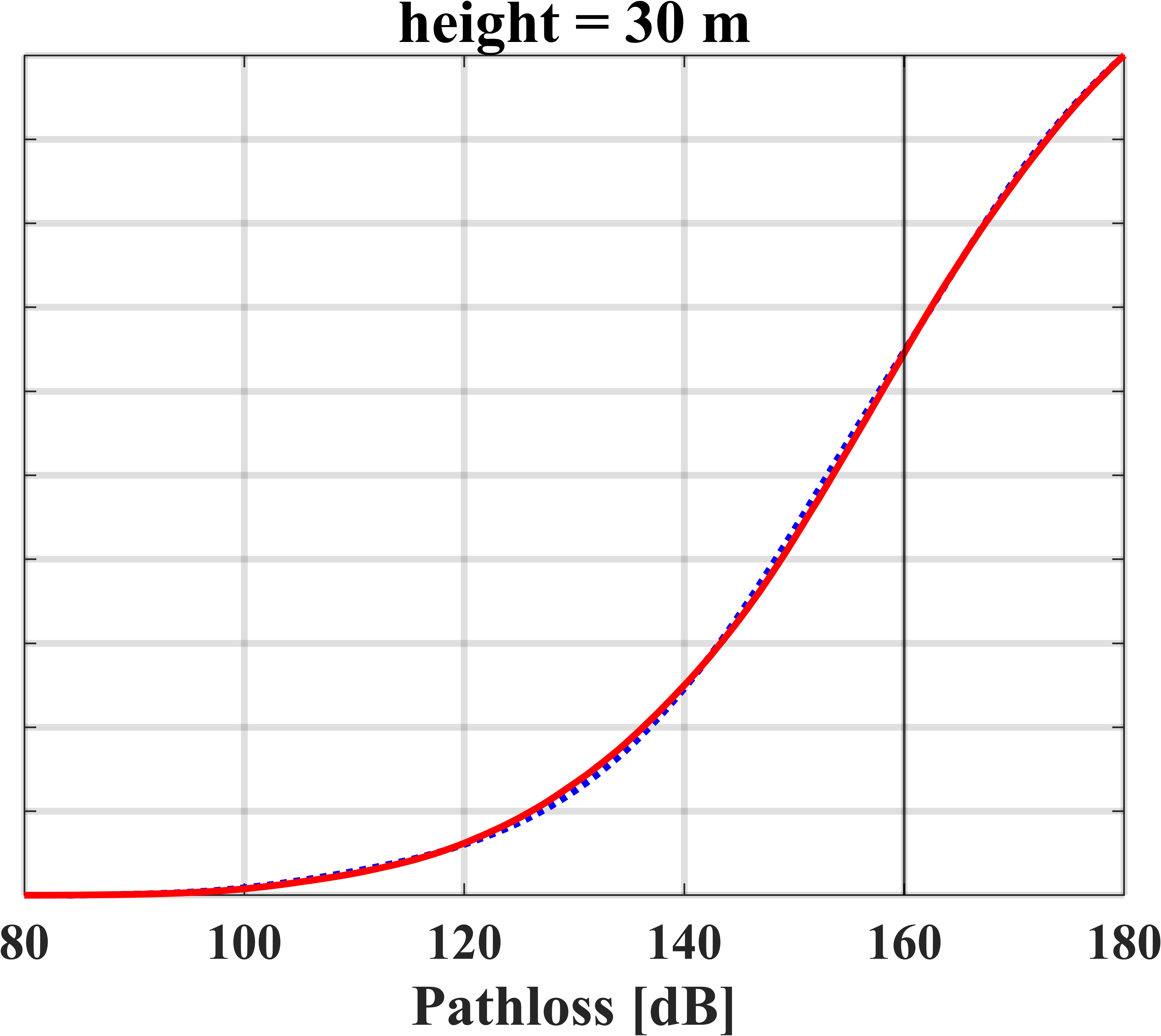} \label{fig:pl_30}}
\subfloat[][]
{\includegraphics[width =0.38\columnwidth]{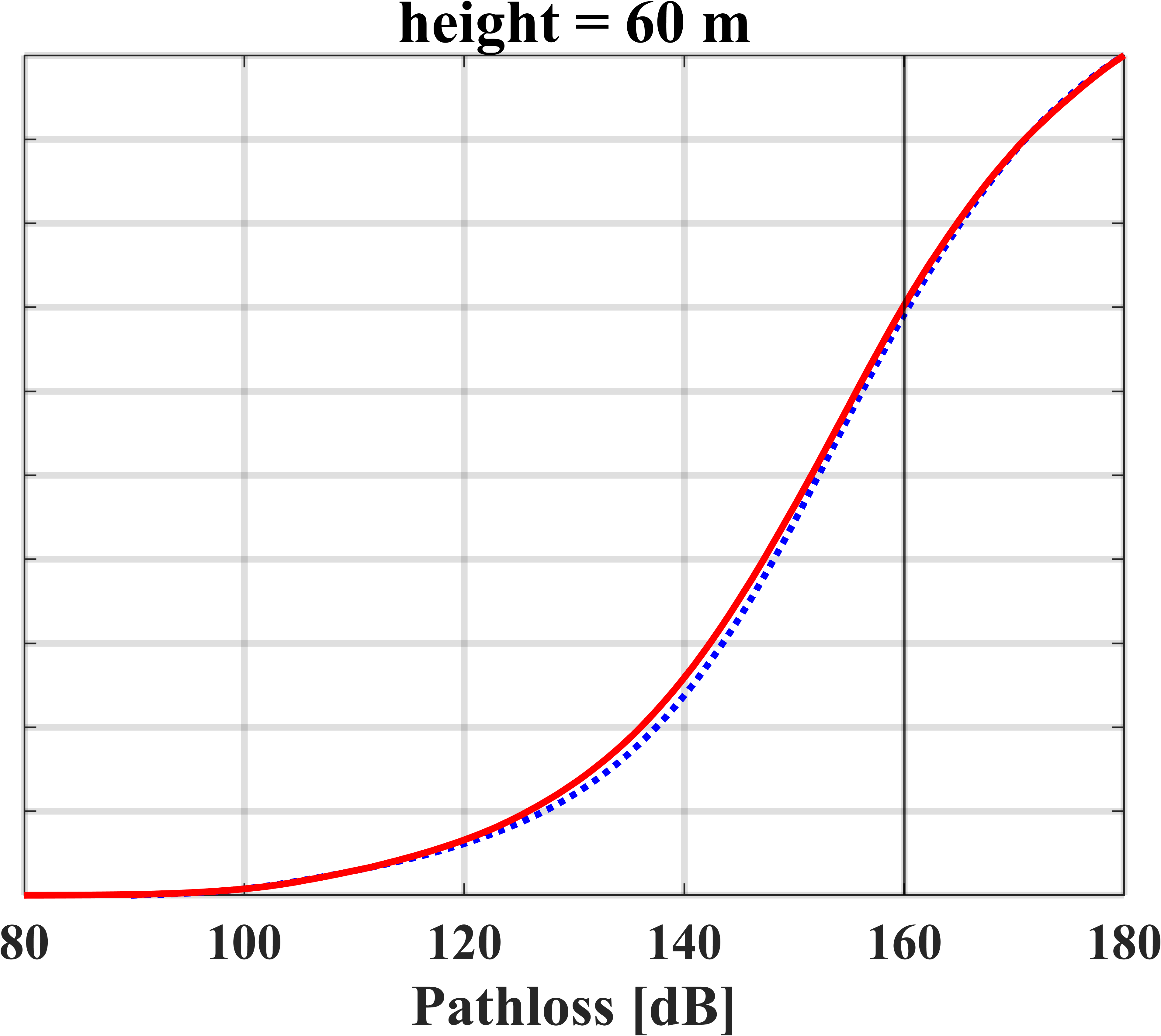} \label{fig:pl_60}}
\subfloat[][]
{\includegraphics[width =0.38\columnwidth]{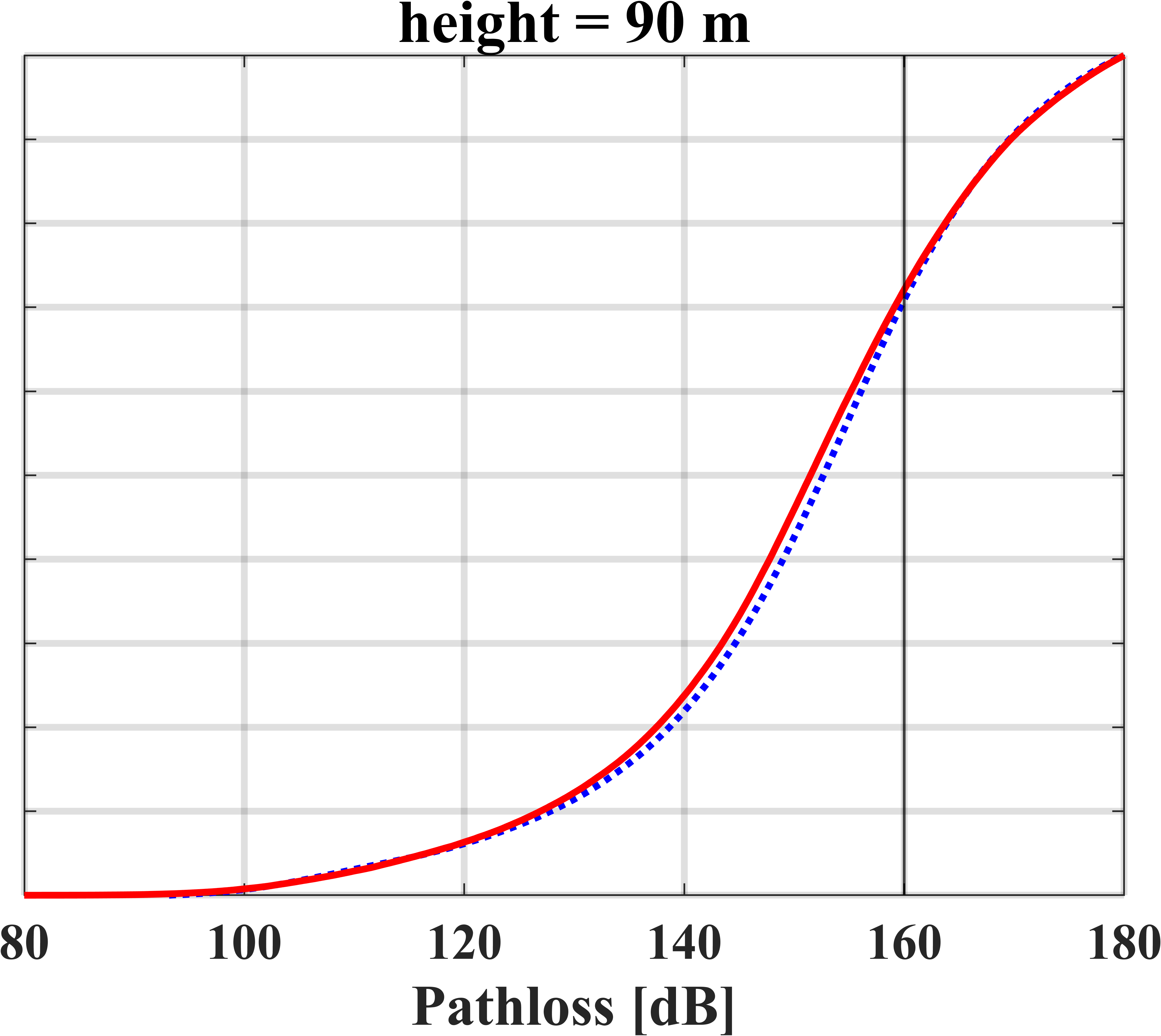} \label{fig:pl_90}}
\subfloat[][]
{\includegraphics[width =0.38\columnwidth]{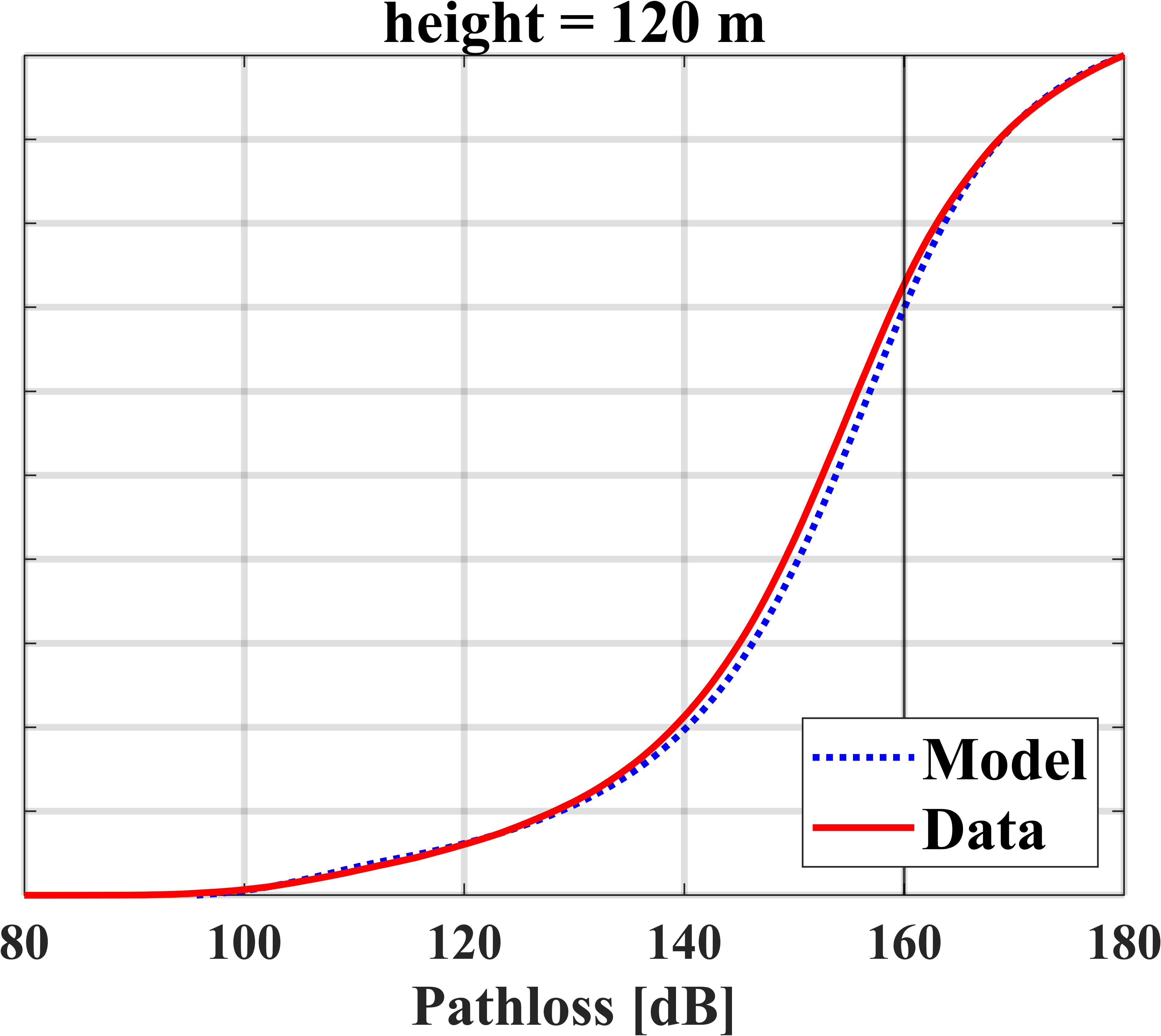} \label{fig:pl_120}}
\caption{CDFs of pathloss for (a) $\SI{1.6}{m}$, (b) $\SI{30}{m}$, (c) $\SI{60}{m}$, (d) $\SI{90}{m}$, and (e) $\SI{120}{m}$}
\label{fig:pl_cdf}
\end{figure*}

\begin{figure*}[!t]
\centering
\subfloat[][]
{\includegraphics[width =0.425\columnwidth]{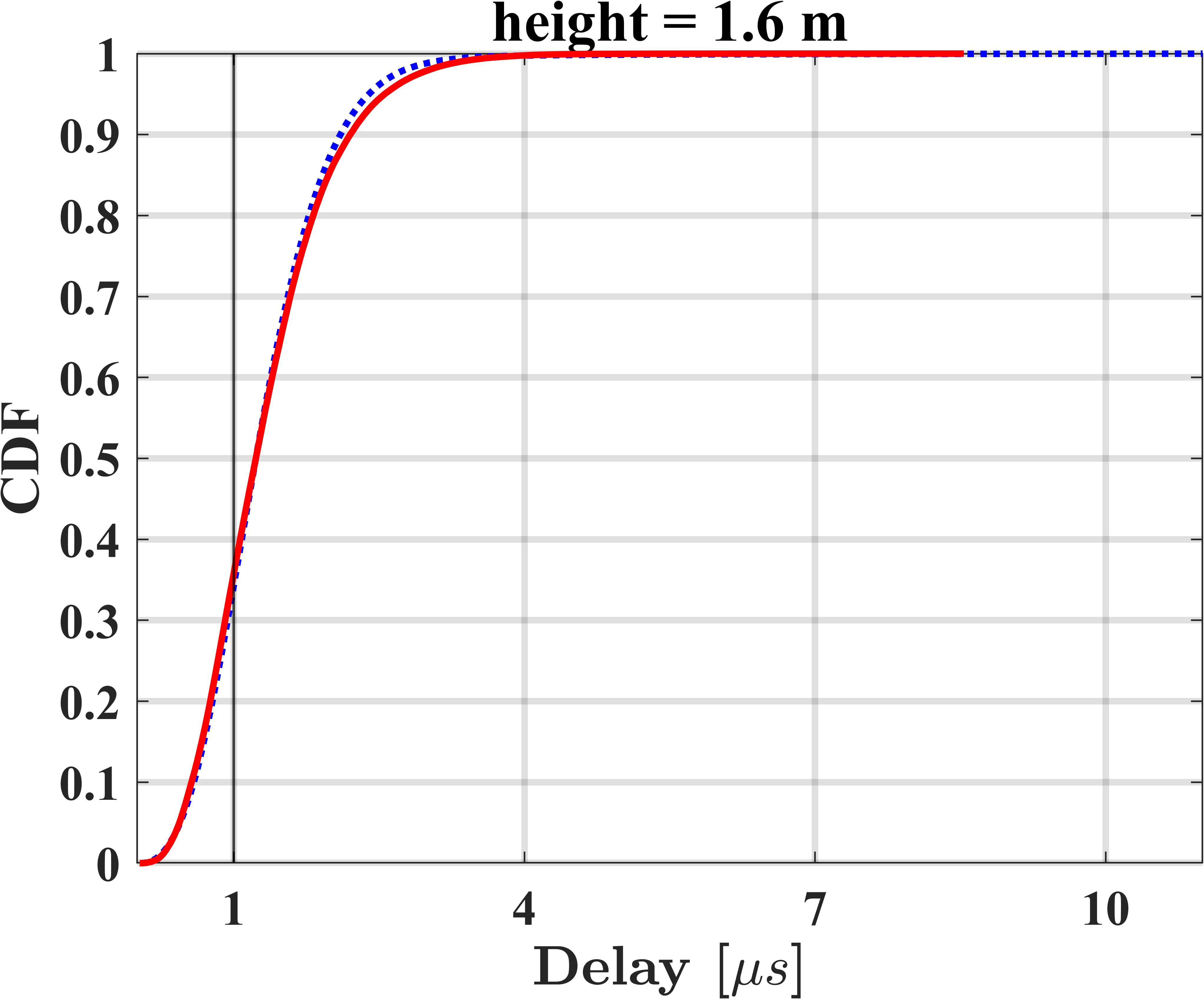} \label{fig:dly_1_6}}
\subfloat[][]
{\includegraphics[width =0.38\columnwidth]{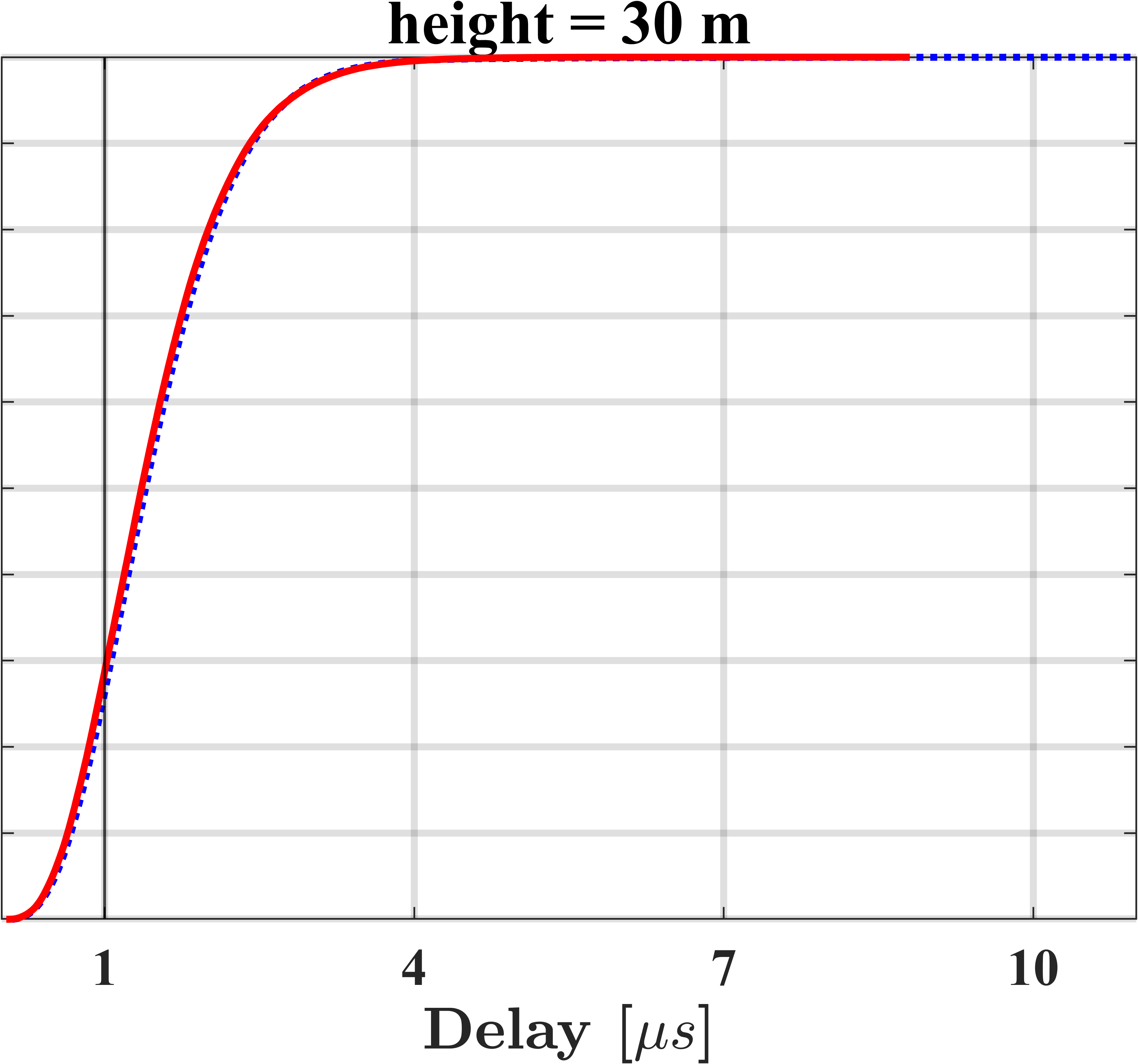} \label{fig:dly_30}}
\subfloat[][]
{\includegraphics[width =0.38\columnwidth]{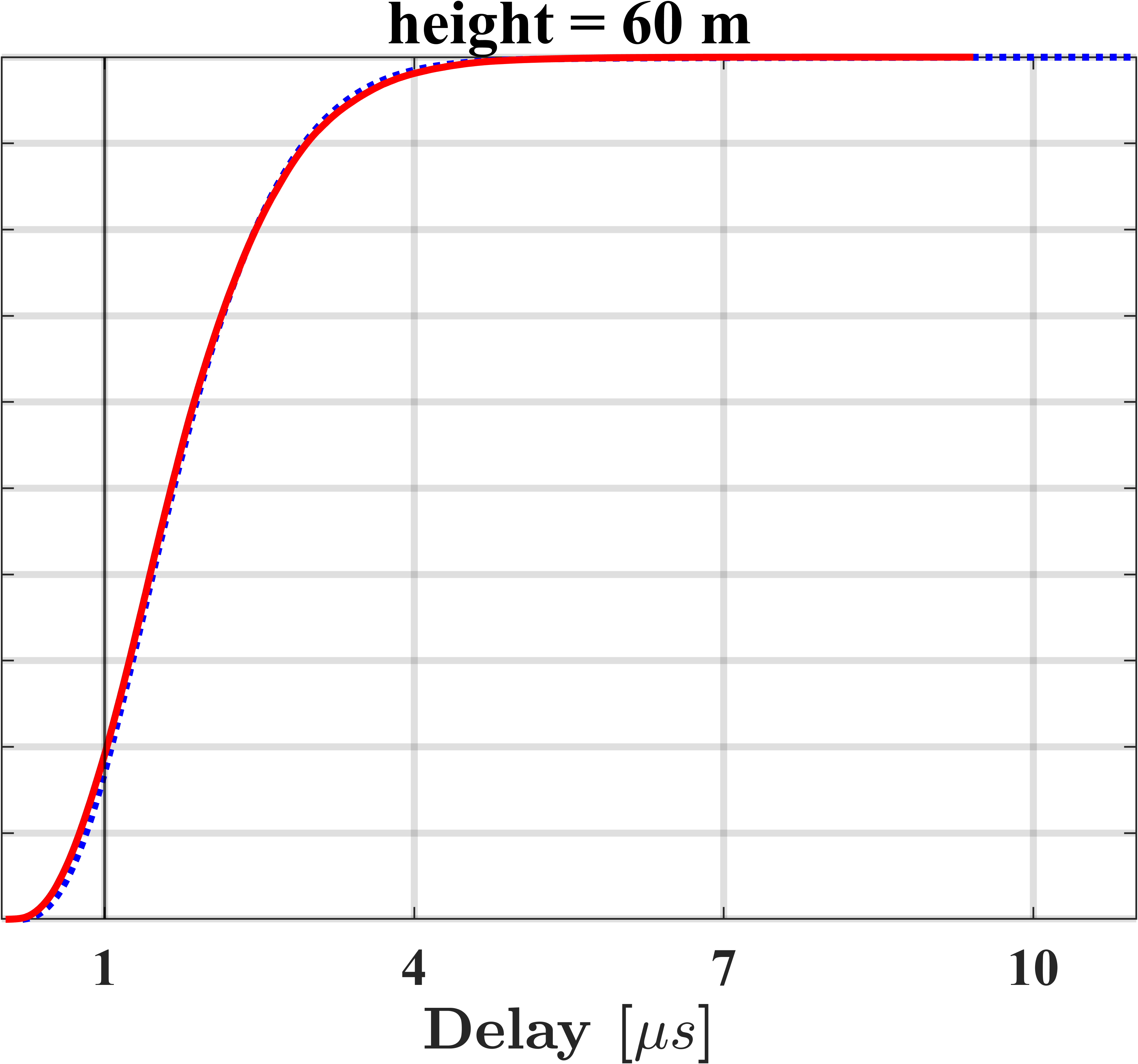} \label{fig:dly_60}}
\subfloat[][]
{\includegraphics[width =0.38\columnwidth]{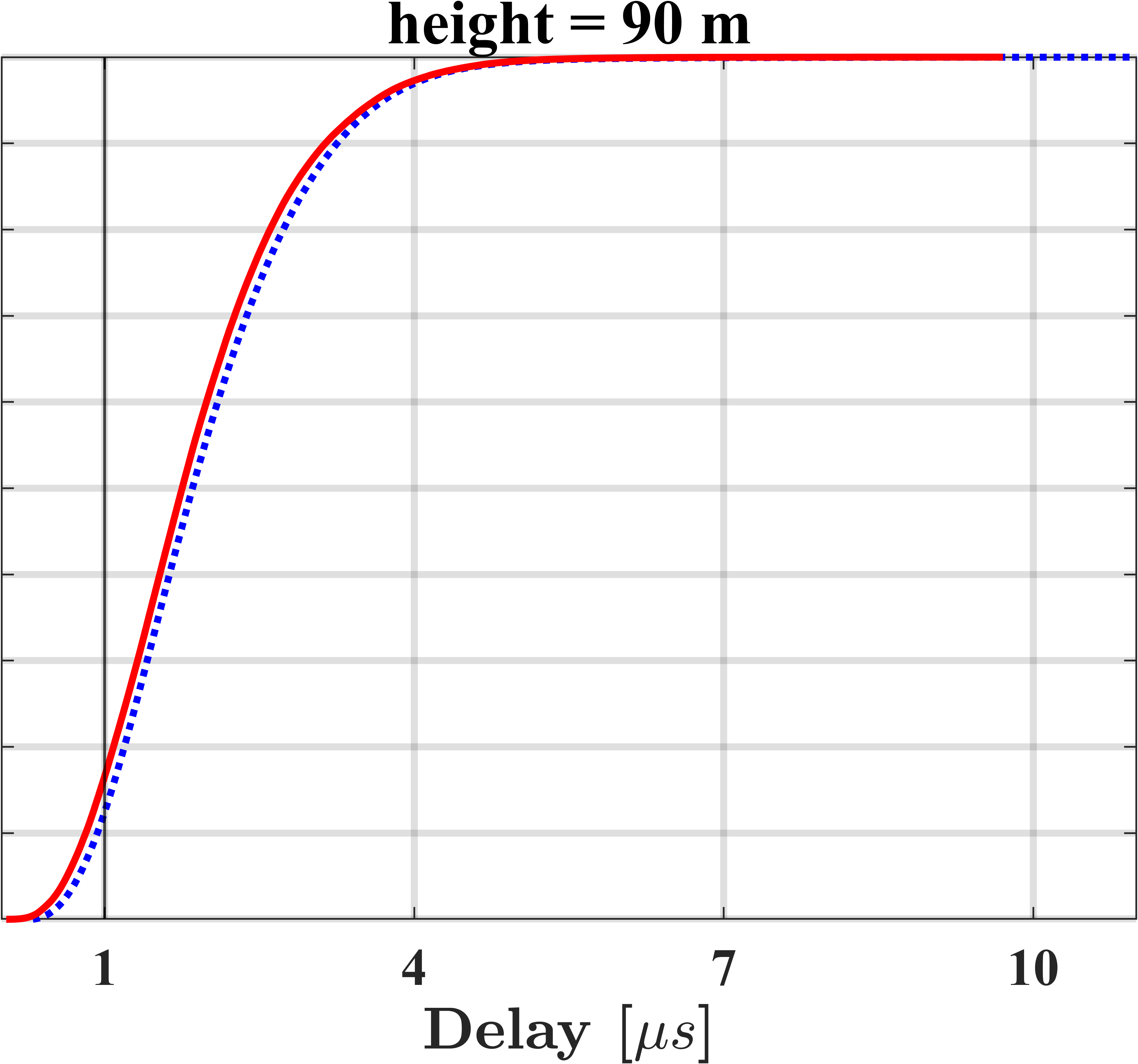} \label{fig:dly_90}}
\subfloat[][]
{\includegraphics[width =0.38\columnwidth]{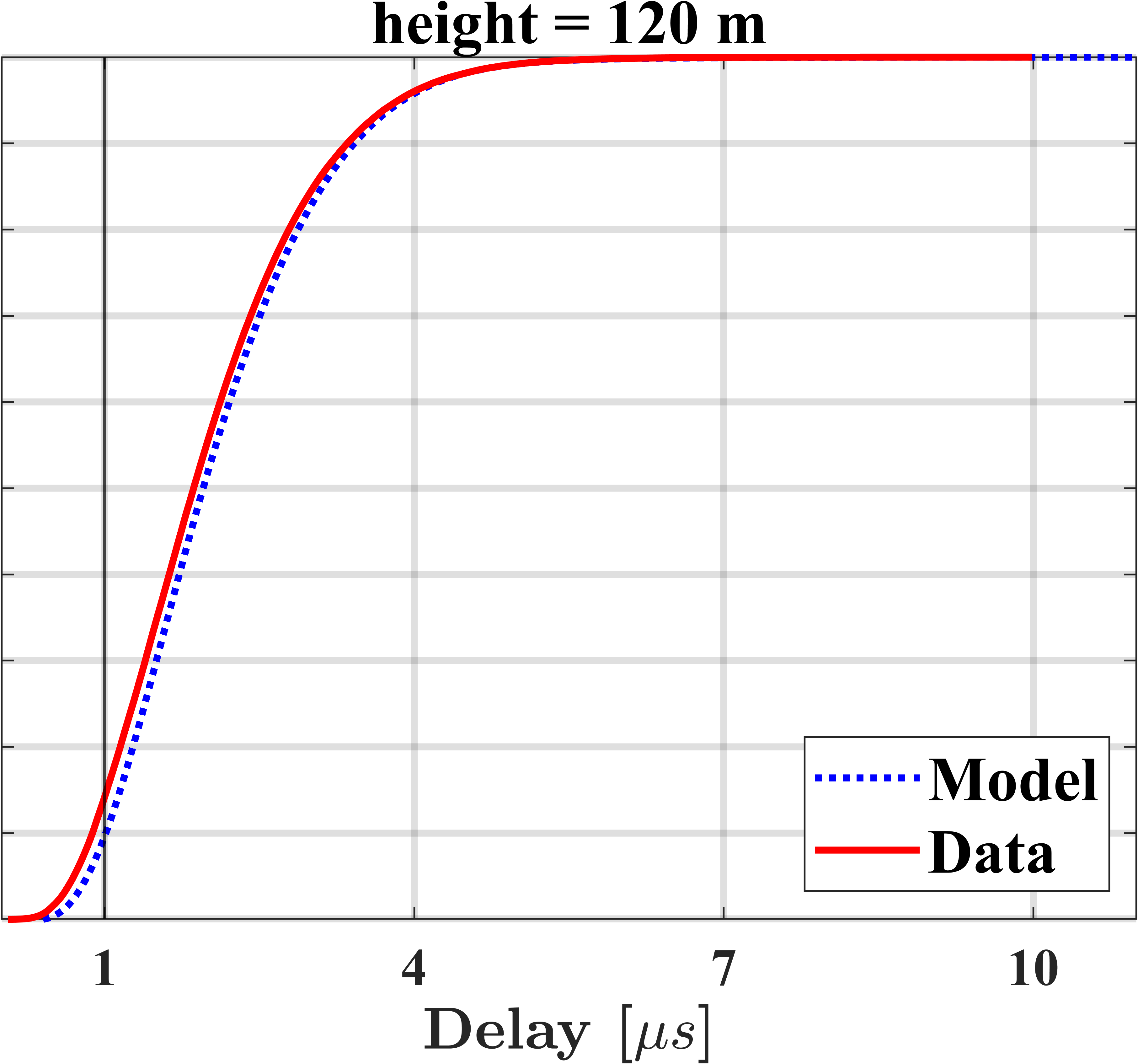} \label{fig:dly_120}}
\caption{ CDFs of propagation delay for (a) $\SI{1.6}{m}$, (b) $\SI{30}{m}$, (c) $\SI{60}{m}$, (d) $\SI{90}{m}$, and (e) $\SI{120}{m}$}
\label{fig:dly_cdf}
\end{figure*}
\section{MODEL TRAINING AND DATA RECONSTRUCTION}
\label{sec:6_model_train_recover}
We confirmed that within $20$ epochs of training, the channel images generated from the model become similar to the original ones. 
The hyperparameters for training are listed in Table~\ref{tab:gan_hyperparams}.
Fig.~\ref{fig:critic_loss} shows that the negative critic loss during model training converges to zero. Considering the meaning of critic loss by Eq.~\ref{eq:critic_loss}, we observe that the generator minimizes the EM distance $W\left(\mathbb{P}_\r, \mathbb{P}_\btheta\right)$ as the loss converges. 
Thus, we stopped training the model when the negative critic loss is close to $0$.

After training the model, the output matrices $\Tilde{\boldsymbol{D}}_{\rm image} \in \mathbb{R}^{3 \times 64 \times 50}$ of WGAN-GP are processed to recover the original range of values, following reversely
all the steps described in Section \ref{sec:data_processing}.
The detailed data reconstruction procedures are followings: 

\begin{itemize}
    \item First, the size of the output matrices should be reduced using an appropriate sampling method.
    \begin{equation*}
    \Tilde{\boldsymbol{D}}_{\rm image} \in \mathbb{R}^{3 \times 64 \times 50} \rightarrow \Tilde{\boldsymbol{D}} \in \mathbb{R}^{8 \times 25}
    \end{equation*}
     In this work, we simply sample the first element of the repeated area: $[\Tilde{\boldsymbol{D}}]_{i, j} = \left[\Tilde{\boldsymbol{D}}_{\rm image}\right]_{0, 8\times i, 2 \times j}$ where $i = 0,1, \cdots, 7$ and $j = 0, 1 \cdots, 24$.
    \item After reducing the size of the model outputs, all elements are recovered by \emph{inverse} Min-Max scaling. In particular, to recover pathloss and delay values, FSPL and minimum delay values computed using Eq.~\ref{eq:los_pl} and Eq.~\ref{eq:dly} are added to pathloss and delay values after applying the inverse Min-Max scaling.
    \item To determine the state of the link, we take the average of the last row of the output matrix $\Tilde{\boldsymbol{D}} \in \mathbb{R}^{8 \times 25}$. If the average value of the last row is positive, the link state is classified as LOS; otherwise, it is classified as NLOS. 
    \item \textcolor{black}{Since the LOS path is geometrically determined, all channel parameters of the first arrival path must be \emph{deterministic}, whereas the trained model produces inherently stochastic outputs. Therefore, whenever the link state in $\Tilde{\boldsymbol{D}}$ indicates LOS, the first column of $\Tilde{\boldsymbol{D}}$ is replaced by the deterministic values obtained from Eq.~\ref{eq:los_pl}--\ref{eq:los_ps} based on the TX and RX coordinates, consistent with standard channel modeling methodologies such as 3GPP TR 38.901~\cite{3GPP38901}}.
    \item Finally, virtual paths are identified and removed when the pathloss values exceed the outage pathloss threshold ($\SI{180}{dB}$). 
\end{itemize}
In this way, the proposed model stochastically generates a varying number of multipath components conditioned on the input parameters.
\section{EVALUATION ON TRAINED MODEL}
\label{sec:7_model_eval}
In this section, we evaluate the channel statistics captured by the WGAN-GP by comparing its outputs with the original data. 
\textcolor{black}{
\subsection{EM Distance Evaluation}  }
\begin{table}[!t]
\centering
\caption{EM Distance}
\label{tab:em_distance}

\renewcommand{\arraystretch}{1.15}
\setlength{\tabcolsep}{4.5pt}  
\begin{tabular}{cccccccc}
\toprule
 RX height &
$\boldsymbol{p}$&
$\boldsymbol{\tau}$ &
$\boldsymbol{\phi}^{\text{tx}}$&
$\boldsymbol{\theta}^{\text{tx}}$ &
$\boldsymbol{\phi}^{\text{rx}}$ &
$\boldsymbol{\theta}^{\text{rx}}$ &  
$\boldsymbol{\psi}$  \\
\midrule

1.6 m  &0.015 &0.008 &0.018 &0.013 &0.015 &0.006 &0.010 \\
30 m  &0.017 &0.013 &0.019 &0.012 &0.017 &0.006 &0.009 \\
60 m    &0.020 &0.016 &0.024 &0.017 &0.022 &0.013 &0.007 \\
90 m   &0.015 &0.010 &0.024 &0.018 &0.020 &0.010 &0.007 \\
120 m &0.016 &0.011 &0.026 &0.016 &0.024 &0.012 &0.008 \\

\bottomrule
\end{tabular}
\end{table}
\textcolor{black}{
Since the value range of each parameter differs, a direct comparison of the EM distances across parameters would be misleading. To ensure a fair comparison, the EM distance is computed in the normalized domain, i.e., between the model outputs before the data reconstruction process and the original data after normalization, where all parameter values lie within $[-1, 1]$.}
\textcolor{black}{
Table~\ref{tab:em_distance} summarizes the EM distance results. Given that the maximum possible EM distance is
2 in the normalized domain, the measured values are negligible, which implies that the proposed model sufficiently captures the underlying distribution of the original data. While the EM distance quantitatively confirms the overall fidelity, it does not reveal the distributional characteristics of individual channel parameters. The following subsections thus complement this quantitative assessment with visual comparisons of each parameter, including CDF and heatmap analyses.}

\subsection{Pathloss and Delay Distributions}

Fig.~\ref{fig:pl_cdf} shows CDFs of pathloss on all paths over all links. We observe that the CDFs drawn by the data from the model align well with the CDFs from the original data. In particular, the conditionality by heights holds, as we observe that the probability of $\text{pathloss} >\SI{160}{dB}$ decreases for higher altitudes. This tendency can be mainly explained by the higher probability of achieving LOS at higher altitudes.

Similarly, as shown in Fig.~\ref{fig:dly_cdf}, we observe that two CDFs (model and data) of propagation delays on all paths of all links are well matched. In addition, we see the different conditional distributions at each height, as the probability of $\tau<1 \mu s$ decreases at higher altitudes.
This is mainly because, in dense urban areas, the NLOS propagation paths are dominant when the RX heights are lower. 

\subsection{Link State Probabilities}
\begin{figure*}[!t]
\centering
\subfloat[][]
{\includegraphics[width =0.42\columnwidth]{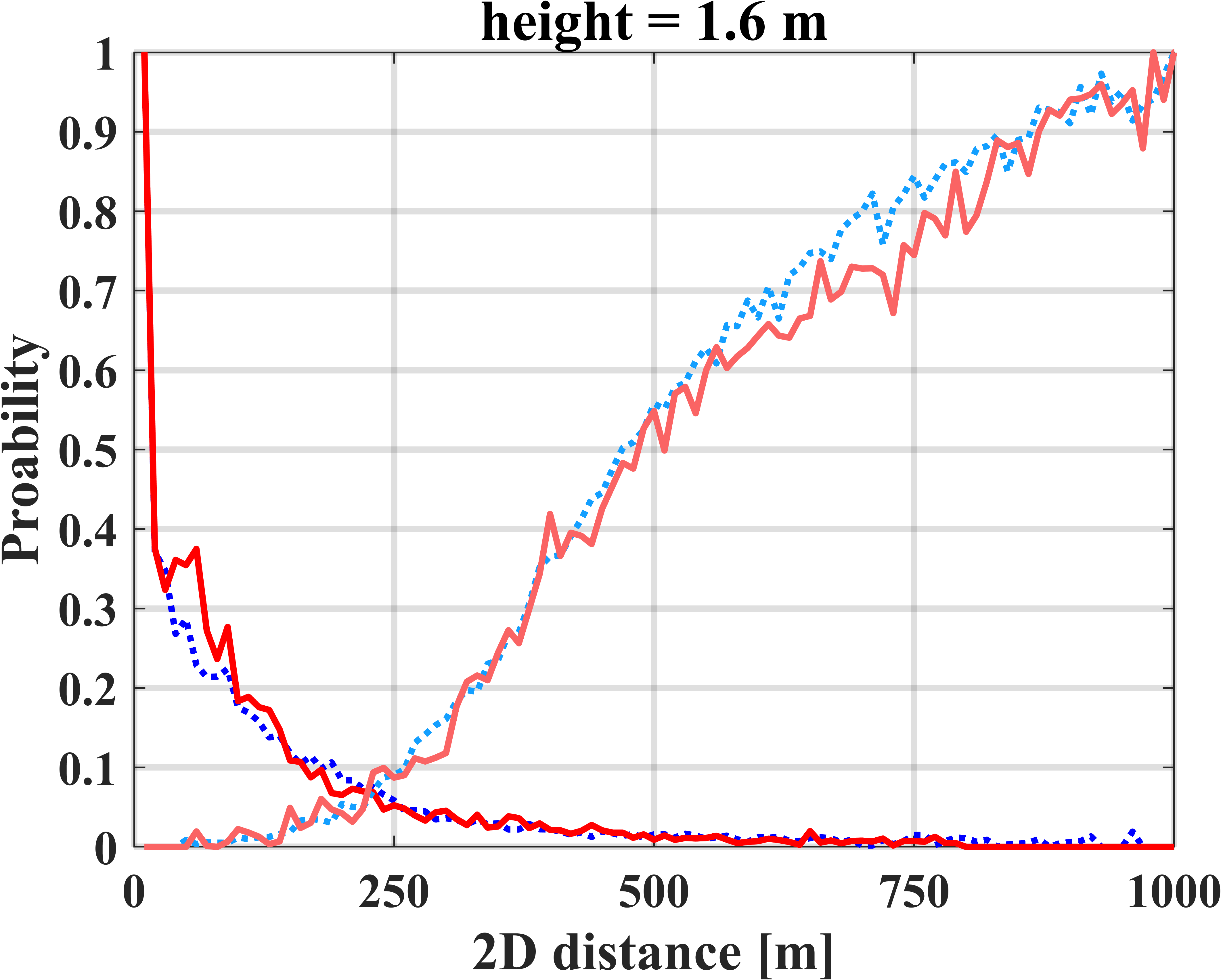} \label{fig:los_prob_1_6}}
\subfloat[][]
{\includegraphics[width =0.38\columnwidth]{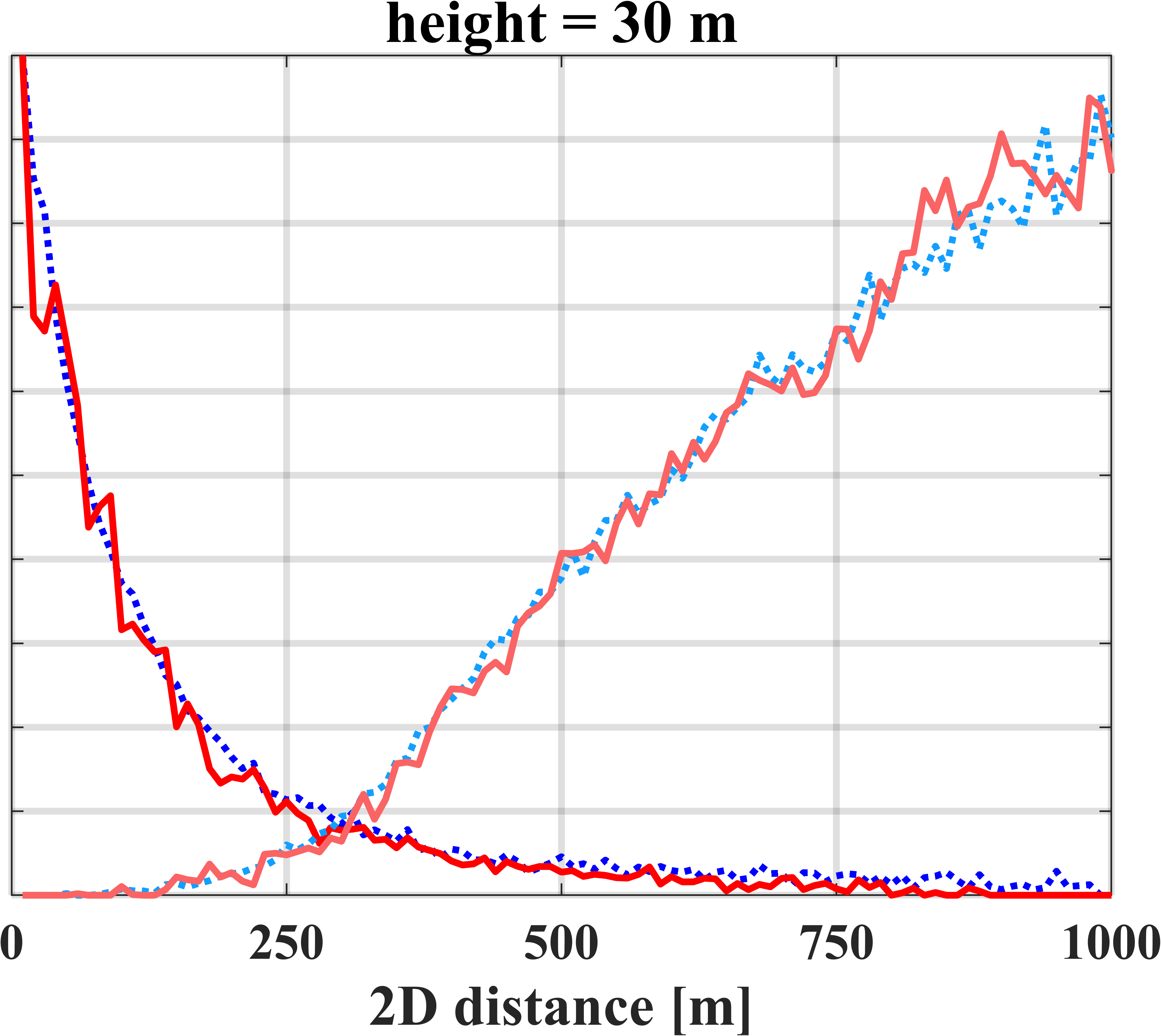} \label{fig:los_prob_30}}
\subfloat[][]
{\includegraphics[width =0.38\columnwidth]{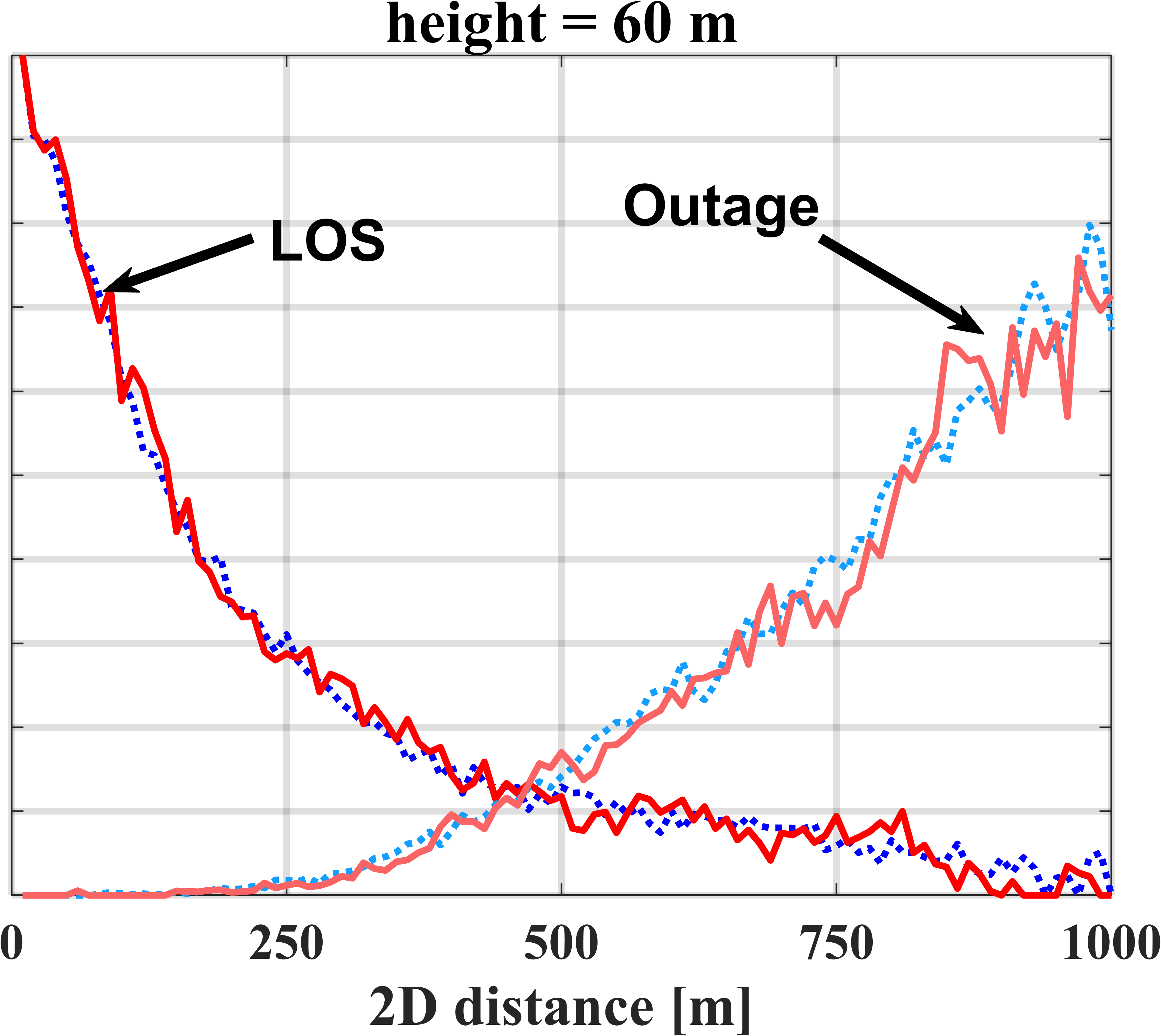} \label{fig:los_prob_60}}
\subfloat[][]
{\includegraphics[width =0.38\columnwidth]{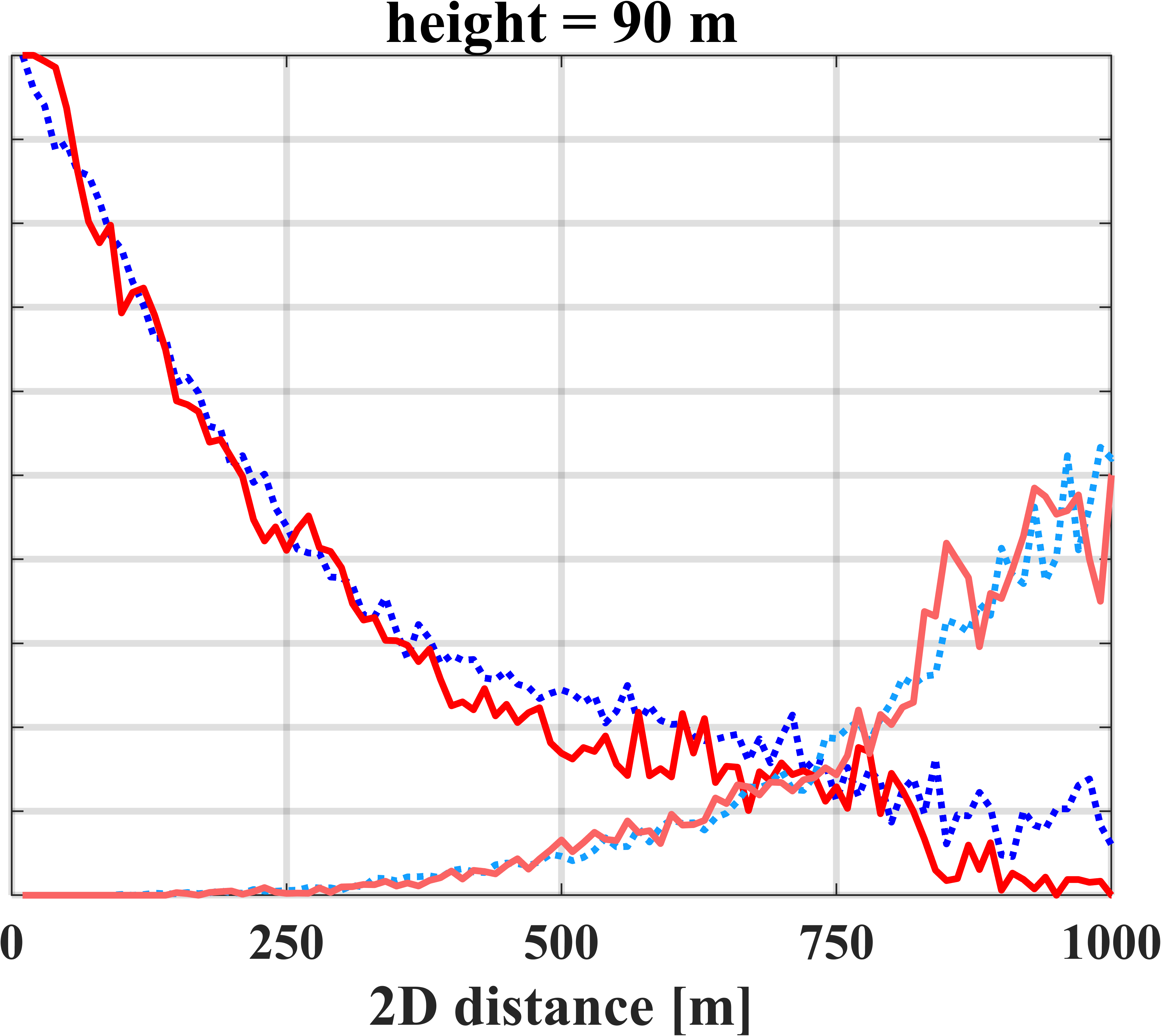} \label{fig:los_prob_90}}
\subfloat[][]
{\includegraphics[width =0.38\columnwidth]{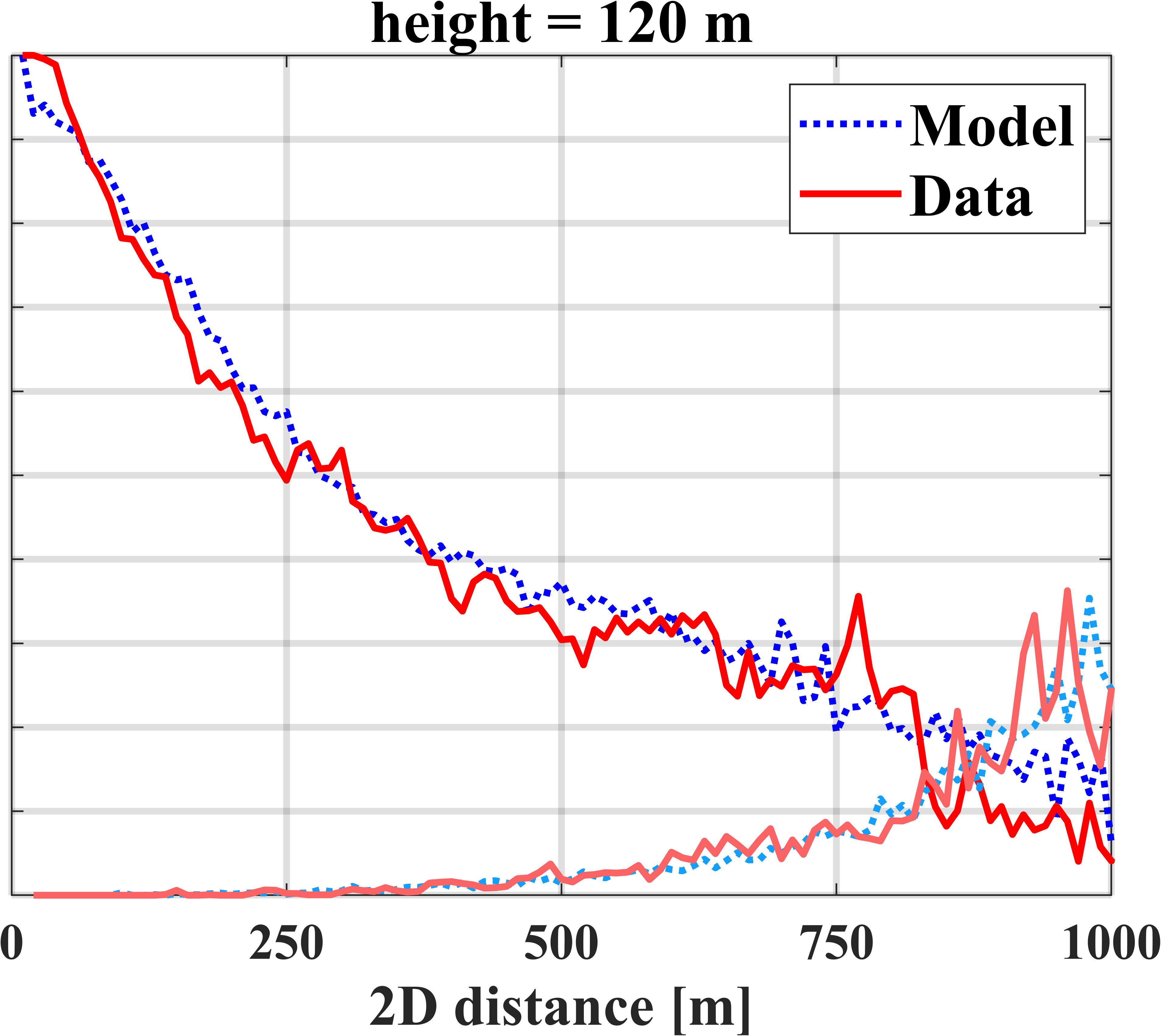} \label{fig:los_prob_120}}
\caption{ Link state probability (LOS and outage) for (a) $\SI{1.6}{m}$, (b) $\SI{30}{m}$, (c) $\SI{60}{m}$, (d) $\SI{90}{m}$, and (e) $\SI{120}{m}$}
\label{fig:los_cdf}
\end{figure*}

Fig.~\ref{fig:los_cdf} shows the probabilities of LOS and link failure at different 2D distances. We define a link outage if all pathloss values of a multipath are greater than $\SI{180}{dB}$.  
As expected, due to the increase in the visibility of RXs at higher altitudes,  the probability of LOS is higher at higher altitudes, whereas the probability of outage is lower. 

In Fig.~\ref{fig:los_cdf}, we can confirm the relationship between the probability of LOS and the RX height. 
In addition, we see that the link state probabilities (LOS and outage) at each height obtained from our model are well matched with the probabilities calculated with the original data.
This result indicates that the trained model effectively captures both conditional constraints: 2D distances and heights.

\begin{figure}[!t]
\centering
\includegraphics[width =0.99\columnwidth]{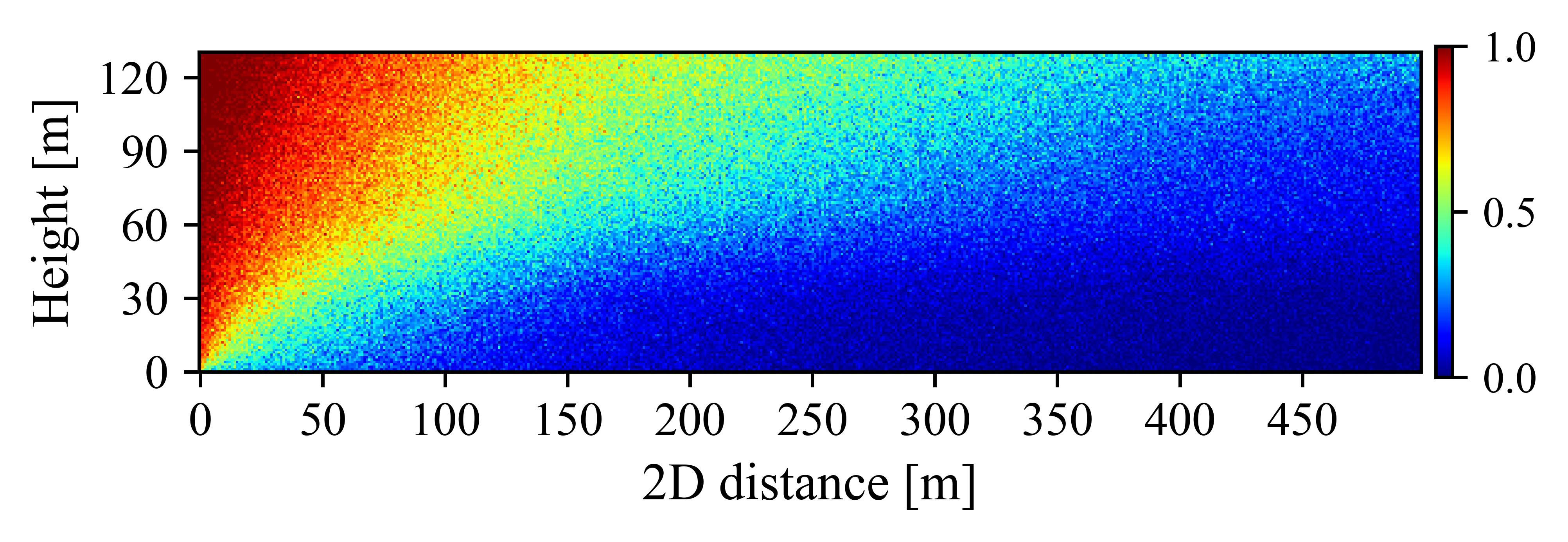} 
\caption{Link state probability by RX height and 2D distance}
\label{fig:los_heatmap}
\end{figure}

Furthermore, Fig.~\ref{fig:los_heatmap} shows the probability of LOS for all 2D distances and RX heights. Although we only used five discrete values of the RX heights, we can observe that the model shows the probable trend of LOS probability over all RX heights. 
This result implies that the trained model performs the statistical interpolation with respect to the RX heights.

\subsection{Distributions of Arrival and Departure Angles}
To examine the angular distributions from the trained model, we take into account the angles relative to the LOS directions. In this case, we can only evaluate the statistics of the ZOD and ZOA under the conditions of 2D distances and heights. We define a random variable, a relative angle to the LOS direction, simply by subtracting the ZOD and ZOA of the LOS, which can be computed using Eq.~\ref{eq:zod} and ~\ref{eq:zoa}. Consequently, the probability density function (PDF) for the relative angle of ZOD conditioned on 2D distance ${\rm dist_{\rm 2d}}$ and height ${\rm h}$ is described as follows:
\begin{align}
    P( \txtheta(&{\rm dist_{\rm 2d}}, {\rm h}) - {\txtheta_{\rm LOS}({\rm dist_{\rm 2d}}, {\rm h})}) \nonumber \\
    &=  P\left( {\txtheta}({\rm dist_{\rm 2d}}, {\rm h}) -  \arctan({\frac{{\rm dist}_{\rm 2d}}{\rm h}})\right)
    \label{eq:pdf_zoda}
\end{align}
In the same way, we define the PDF of ZOAs relative to LOS direction using Eq.~\ref{eq:zoa}. 

\begin{figure}[!t]
\centering
\subfloat[][]
{\includegraphics[width =0.99\columnwidth]{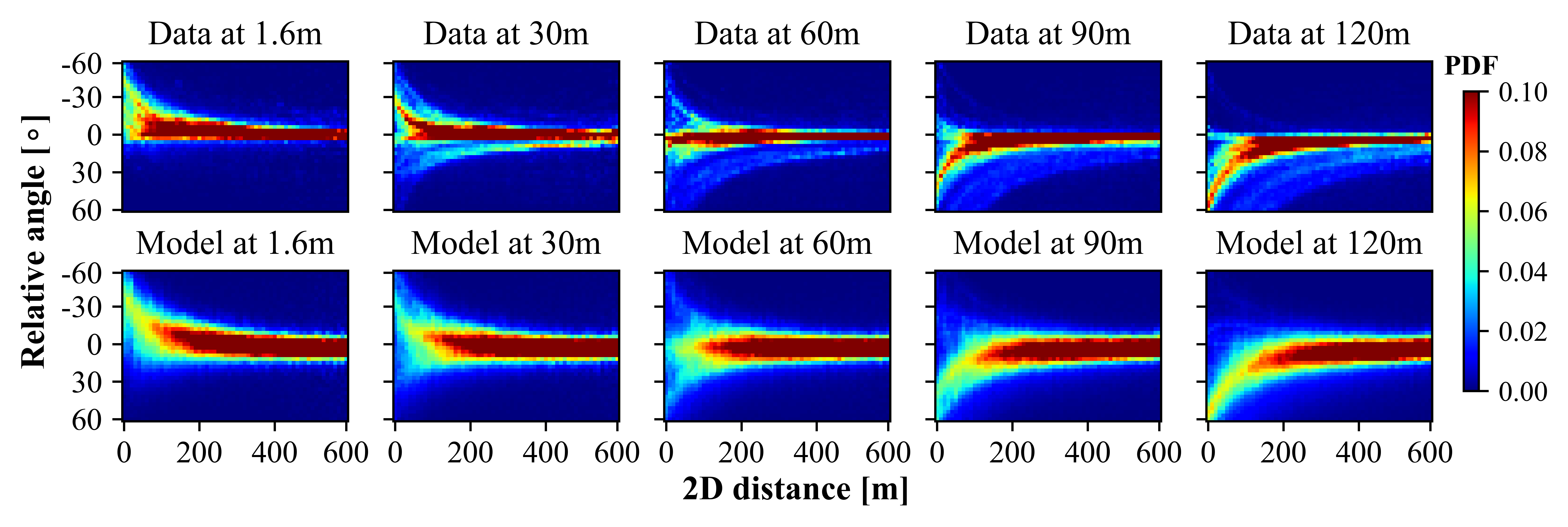} \label{fig:pdf_zod}}
\\
\subfloat[][]{
\includegraphics[width =0.99\columnwidth]{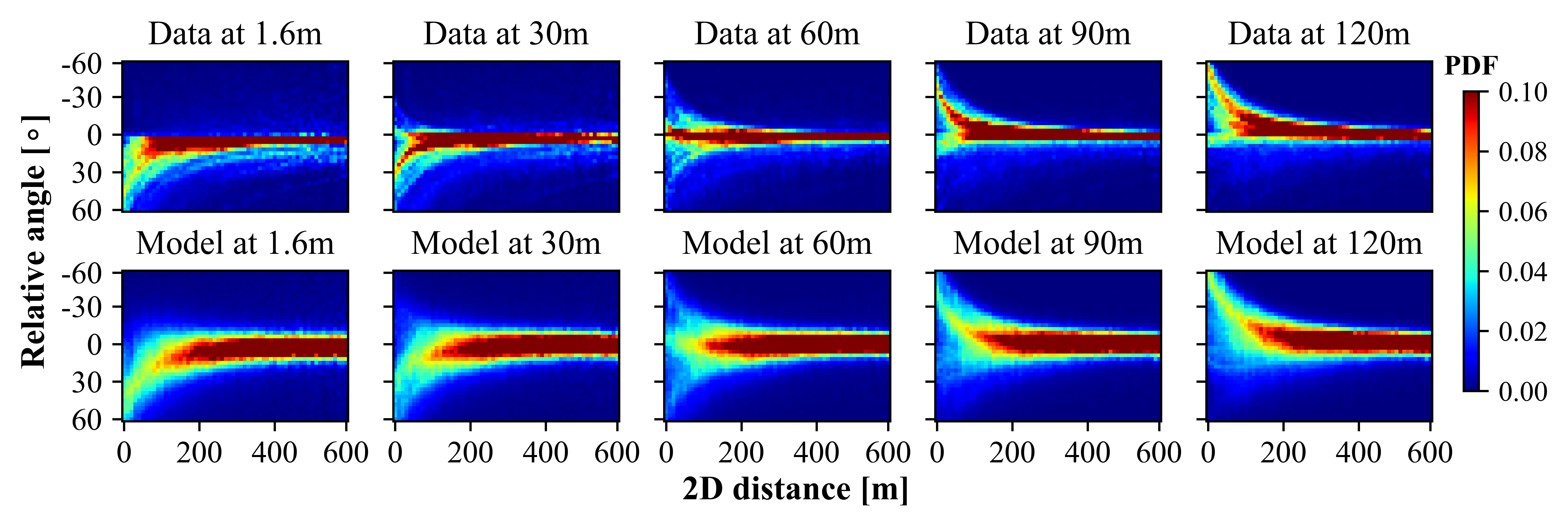}
\label{fig:pdf_zoa}}
\caption{ PDFs of relative (a) ZOD and (b) ZOA  over 2D distances}
\label{fig:pdf_zod_zoa}
\end{figure}
\begin{figure}[!t]
\centering
\includegraphics[width =0.99\columnwidth]{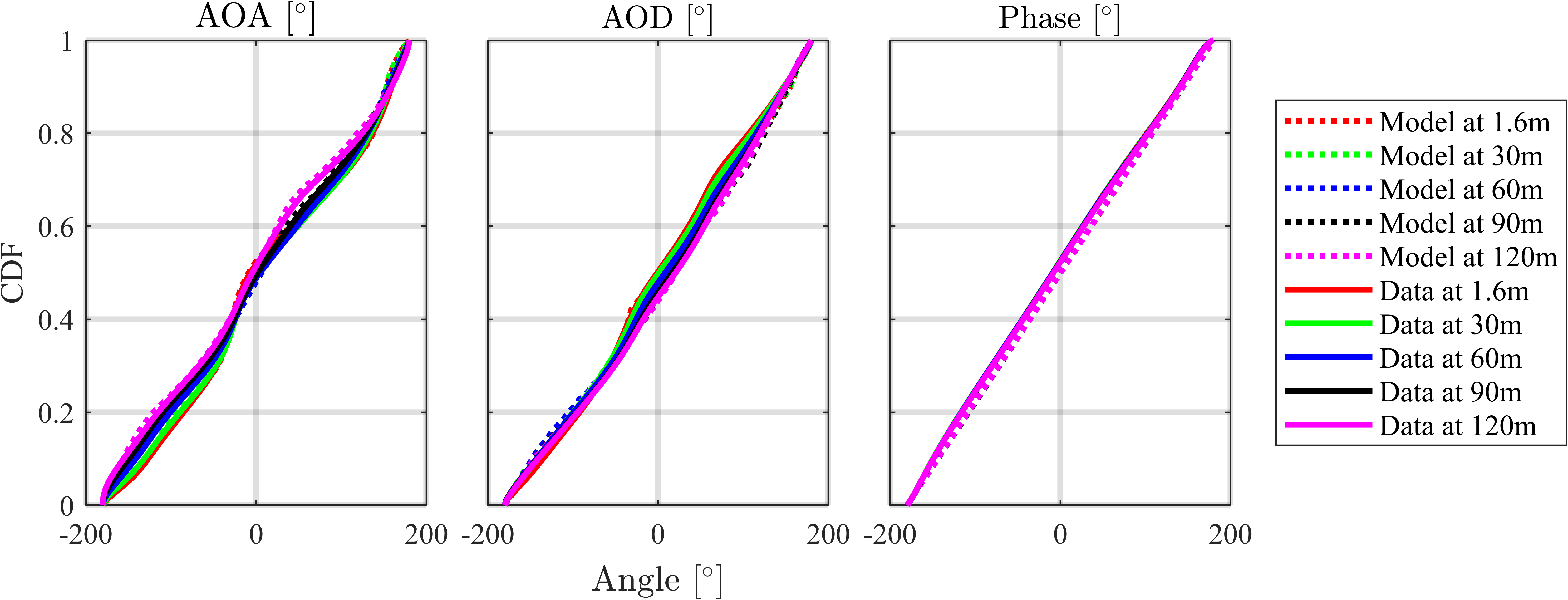} 
\caption{CDFs of AOA, AOD and arrival phase}
\label{fig:azm_phase}
\end{figure}

Fig.~\ref{fig:pdf_zod_zoa} depicts the PDFs of ZOA and ZOD as heatmaps obtained from the original data and outputs of our model at different RX heights. Each column in the heatmaps represents the PDF calculated at the corresponding 2D distance. 
On the one hand, as the 2D distances increase, we observe that ZOA and ZOD have less angular spread as a result of less scattering at a far distance. 
On the other hand, since the local scattering increases at a closer distance, we see a large variation of the relative angles until \SI{200}{m} of the 2D distance.

Specifically, local scattering occurs mostly in downward directions at close distances. Accordingly,  Fig.~\ref{fig:pdf_zod_zoa}(a) and Fig.~\ref{fig:pdf_zod_zoa}(b) illustrate the following interesting results:
\begin{itemize}
    \item On the TX side, we can easily confirm that the LOS ZOD decreases as the RX height increases.   
As shown in Fig.~\ref{fig:pdf_zod_zoa}(a), at \SI{1.6}{m} altitude, we observe a higher probability at negative relative angles (LOS ZODs $>$ ZODs). On the contrary, at \SI{120}{m} altitude, positive relative angles (ZODs $>$ LOS ZODs) have a higher probability as the LOS ZODs decrease. 

\item On the RX side, as opposed to the TX case, the LOS ZOA increases as the RX height increases. 
Therefore, in Fig.~\ref{fig:pdf_zod_zoa}(b), the positive relative angles (ZOAs $>$ LOS ZOAs) have higher probabilities at $\SI{1.6}{m}$ altitude, while at $\SI{120}{m}$ altitude, the probability of negative relative angles (LOS ZOAs $>$ ZOAs) is higher.
\end{itemize}

Since the conditional variables are 2D distances and heights, azimuth angles (AOA, AOD) and arrival phases follow the \emph{uniform} distribution.
Note that for AOA and AOD statistics, Eq.~\ref{eq:pdf_zoda} is not applicable due to the limited conditions and, therefore, it is not possible to show heatmaps of PDFs such as Fig.~\ref{fig:pdf_zod_zoa}(a) and Fig.~\ref{fig:pdf_zod_zoa}(b).
In other words, to calculate LOS azimuth angles, Eq.~\ref{eq:aod} is not applicable due to  
\begin{equation}
 P (\txphi |\text{dist}_{\text{2d}}, \text{h}) \neq   P ( \txphi |\text{dx}, \text{dy}, \text{h}) \nonumber
\end{equation}
where $\rm dx = x_{rx} - x_{tx}$ and $\rm dy = y_{rx} - y_{tx}$.

As shown in Fig.~\ref{fig:azm_phase},  the distributions of AOA, AOD, and arrival phase follow \emph{uniform} distribution at each height owing to the use of the 2D distance conditionality.
\begin{figure*}[!b]
\centering
\subfloat[][]
{\includegraphics[width =0.41\columnwidth]{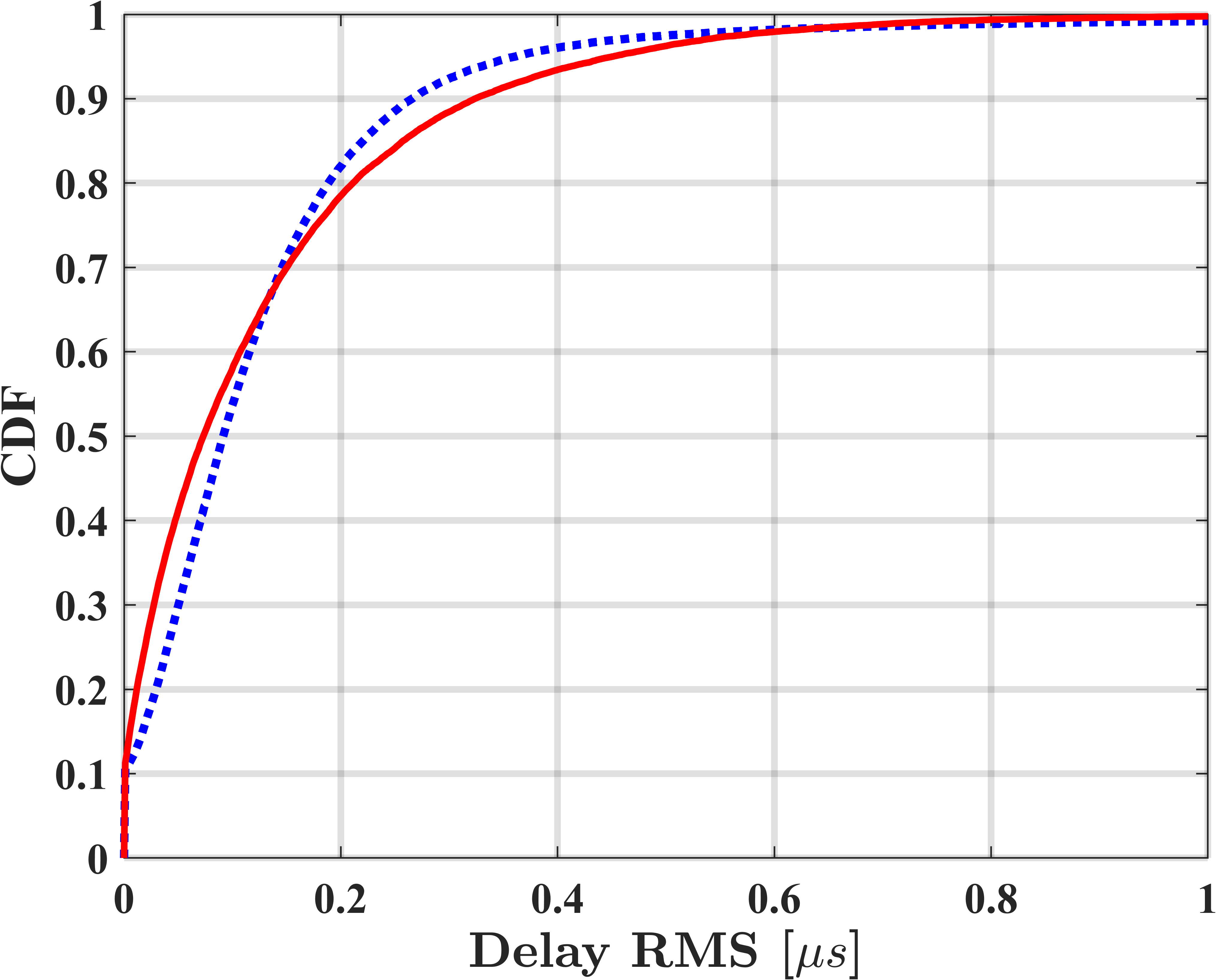} \label{fig:rms_delay}}
\subfloat[][]
{\includegraphics[width =0.38\columnwidth]{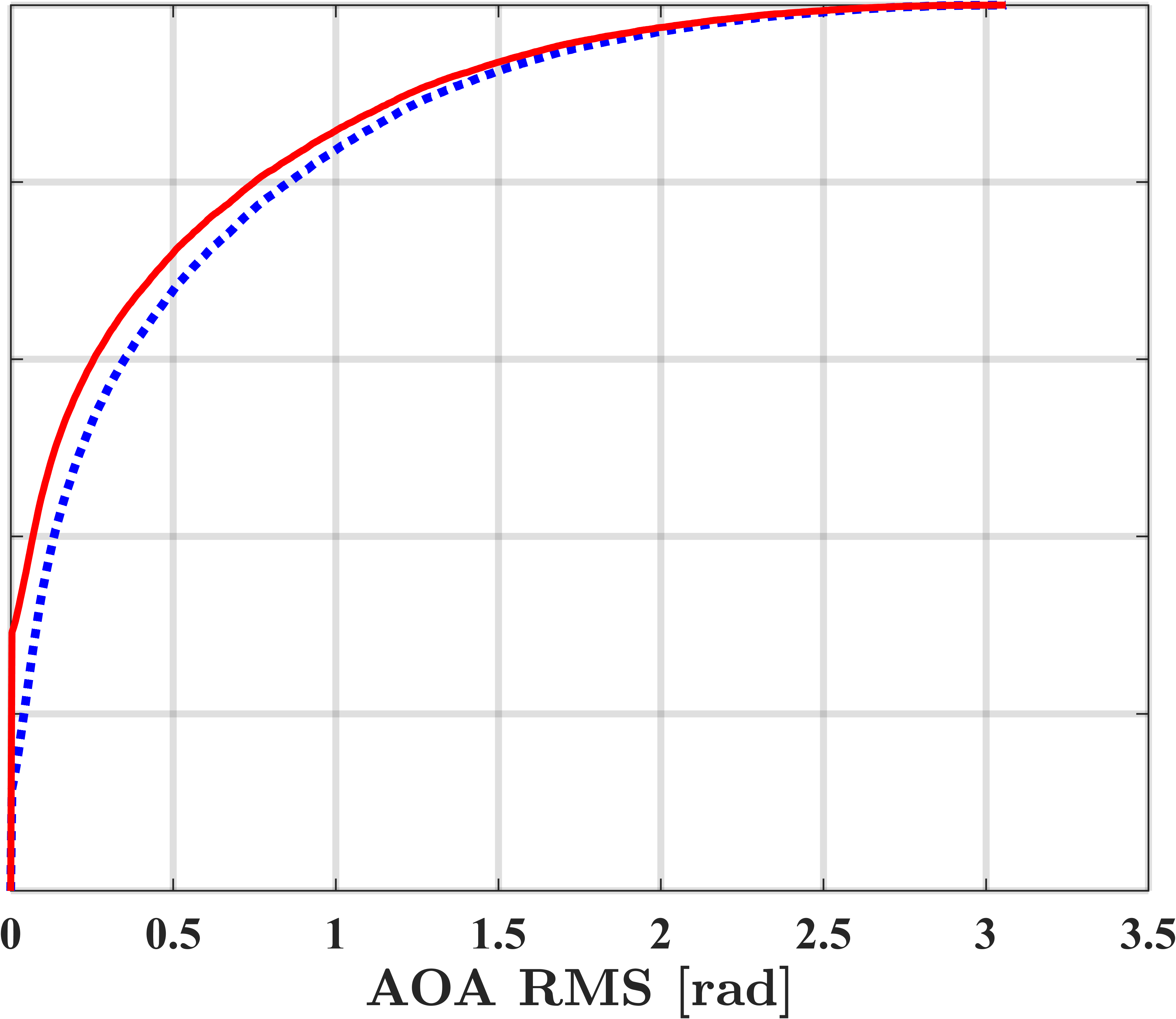} \label{fig:rms_aoa}}
\subfloat[][]
{\includegraphics[width =0.38\columnwidth]{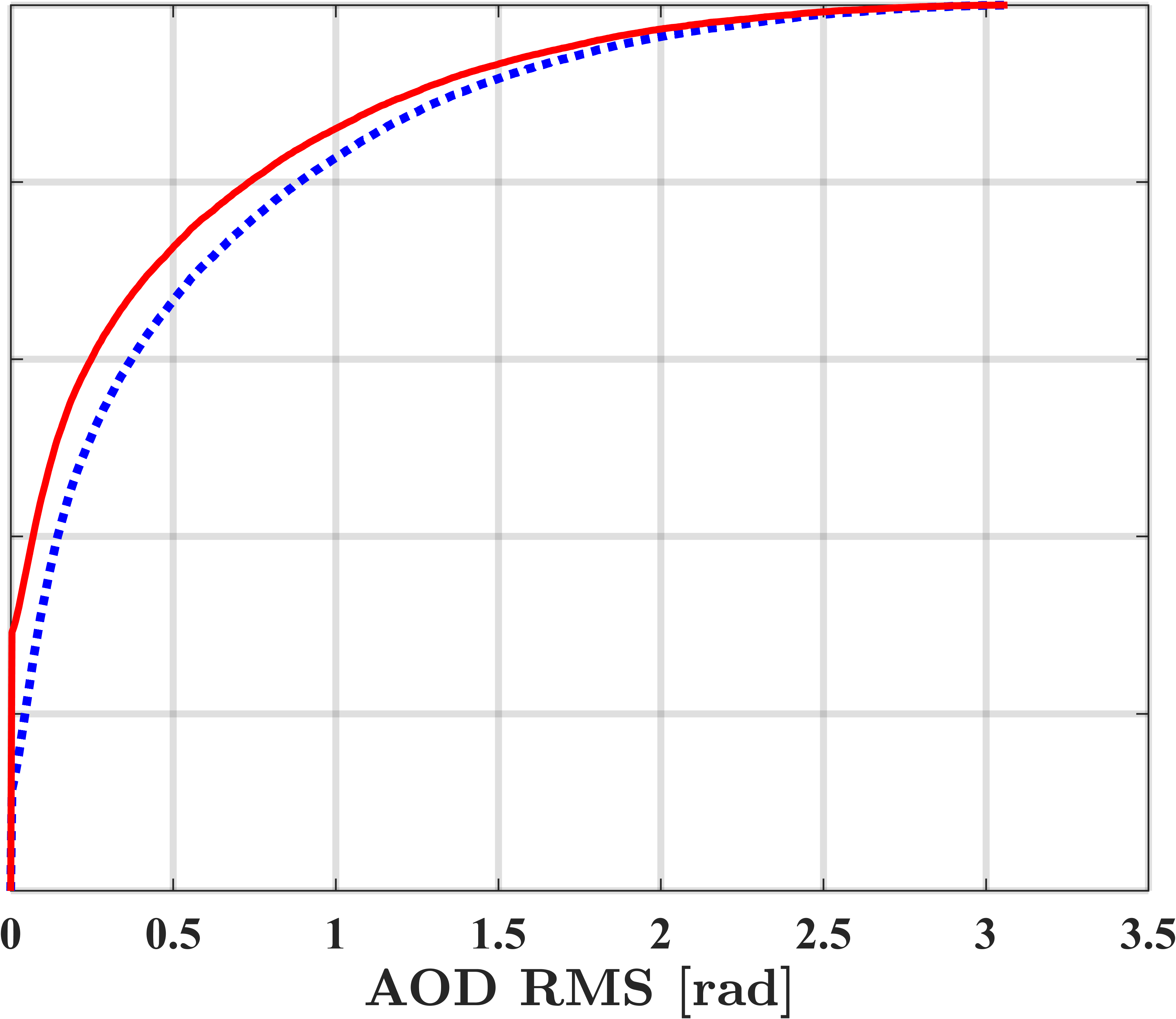} \label{fig:rms_aod}}
\subfloat[][]
{\includegraphics[width =0.38\columnwidth]{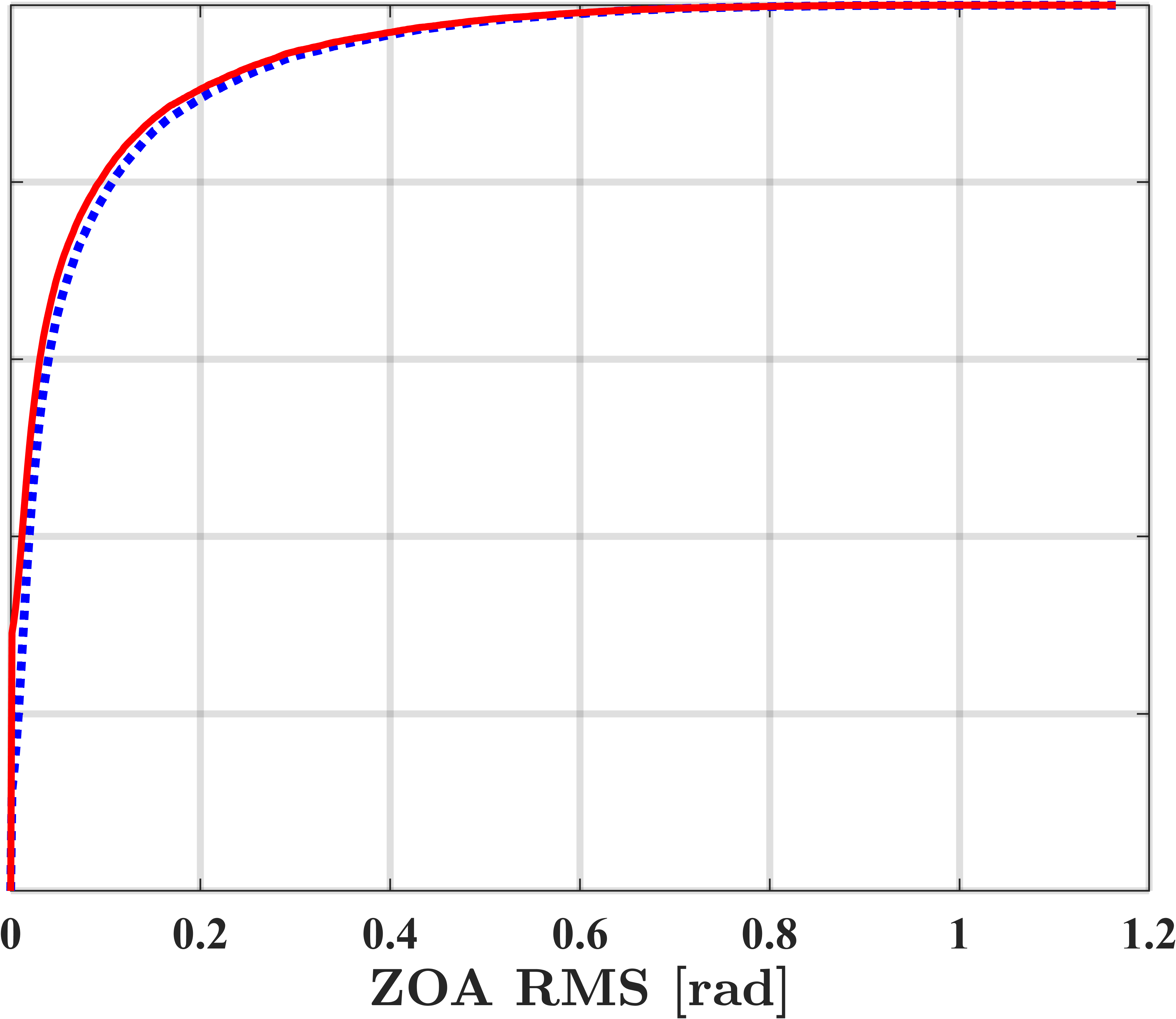} \label{fig:rms_zoa}}
\subfloat[][]
{\includegraphics[width =0.38\columnwidth]{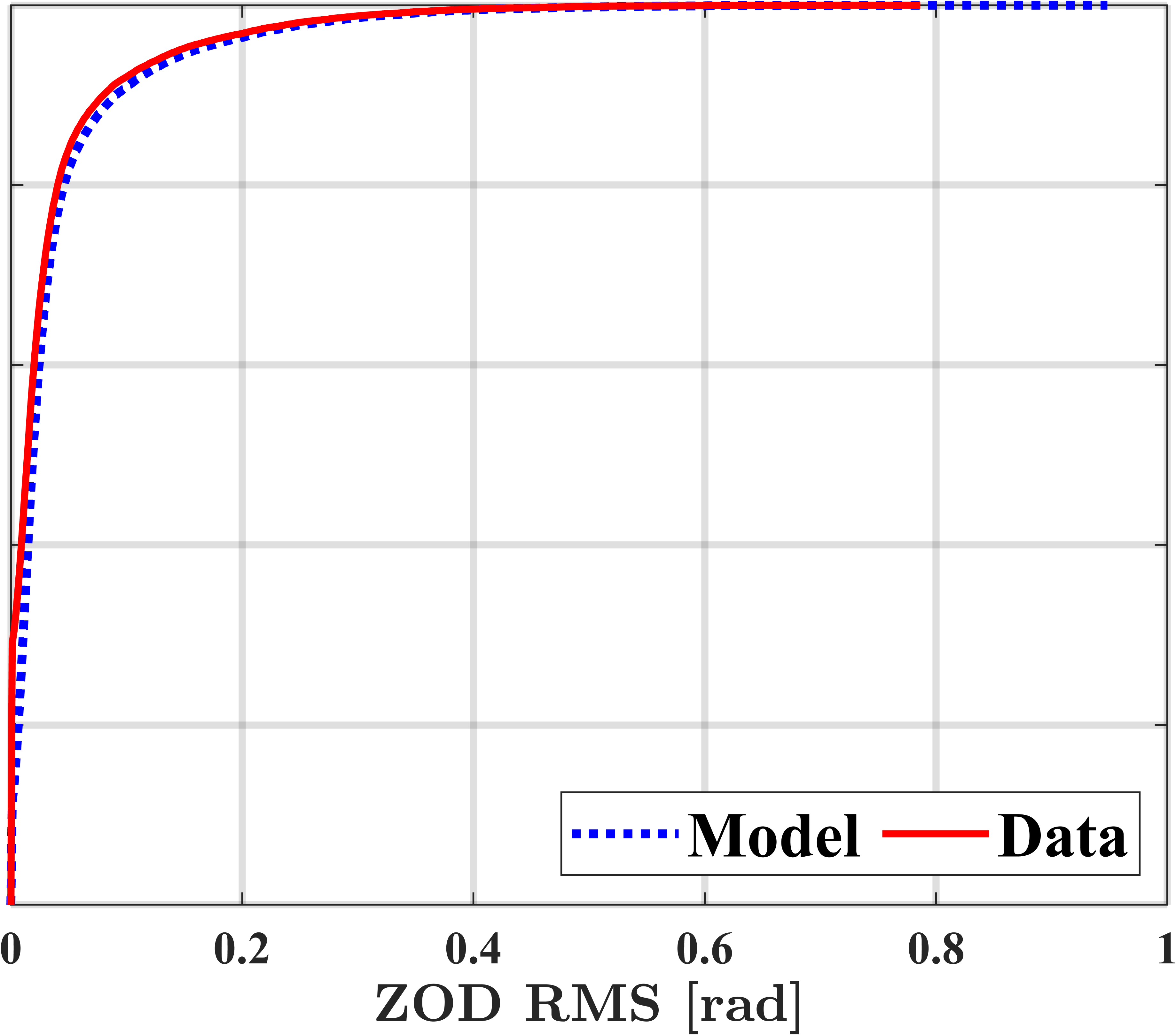} \label{fig:rms_zod}}
\caption{RMS CDFs for (a) delay, (b) AOA, (c) AOD, (d) ZOA, and (e) ZOD}
\label{fig:rms_cdf}
\end{figure*}

\subsection{Distributions of Root Mean Square Values}

To quantify the characteristics of multipath channels, we briefly discuss the root mean square (RMS) spread for the delay, azimuth, and zenith angles, which represents the degree of sparsity or richness of the multipath channels. According to \cite{ITURMS2017}, the general RMS spread values for a multipath feature $\boldsymbol{d}$ are calculated as follows:
\begin{center}
\begin{align*}
    d_{r m s} =\sqrt{\frac{\sum_k\left(d_k-d_m\right)^2 P_k}{\sum_k P_k}},~d_m =\sum_k \frac{d_k P_k}{\sum_k P_k}
\end{align*}
\end{center}
where $P_k$ is path gain, $d_k$ is a channel parameter such as \emph{excess delay}, arrival and departure angles at kth path.
 RMS values are instrumental metrics in dense urban environments where there are many local scatterers. We consider the RMS values of delay, AOA, AOD, ZOA, and ZOD when the RX height is $\SI{1.6}{m}$, as at lower altitudes rich scattering can occur.

The statistical distributions of the RMS values are shown in Fig.~\ref{fig:rms_cdf} as CDF for AOA, AOD, ZOA, ZOD and delay. As shown, we observe that the distributions are very closely aligned with those of the original data. 
 These results imply that the WGAN-GP statistically captures the multipath characteristics that reflect rich or sparse scattering.
 
\subsection{Correlation among Multipath Components}
Capturing the correlations between different multipath components is significant since all the components of a communication link are strongly correlated. However, for simplicity in the analysis and due to the space limit, we only consider the correlations between the \emph{second-strongest path} components $\u = [p_2, \tau_2, \txphi_2, \txtheta_2, \rxphi_2, \rxtheta_2, \psi_2]$.

The normalized correlation matrix, known as Pearson's correlation matrix,  $\boldsymbol{R} \in \mathbb{R}^{7 \times 7} $ for a second-strongest path can be defined as follows:
\begin{align*}
    \boldsymbol{R}_{i j}=\frac{\sum_{k=1}^n\left(\u_{i,k}-\bar{\u}_{i}\right)\left(\u_{j,k}-\bar{\u}_j\right)}{\sqrt{\sum_{k=1}^n\left(\u_{i,k}-\bar{\u}_i\right)^2} \sqrt{\sum_{k=1}^n\left(\u_{j,k}-\bar{\u}_j\right)^2}}
\end{align*}
where $n$ is the number of samples, and the bar symbol $\bar{}$ indicates the average value over the sampled data.

\begin{figure}[!t]
\centering
\includegraphics[width =0.99\columnwidth]{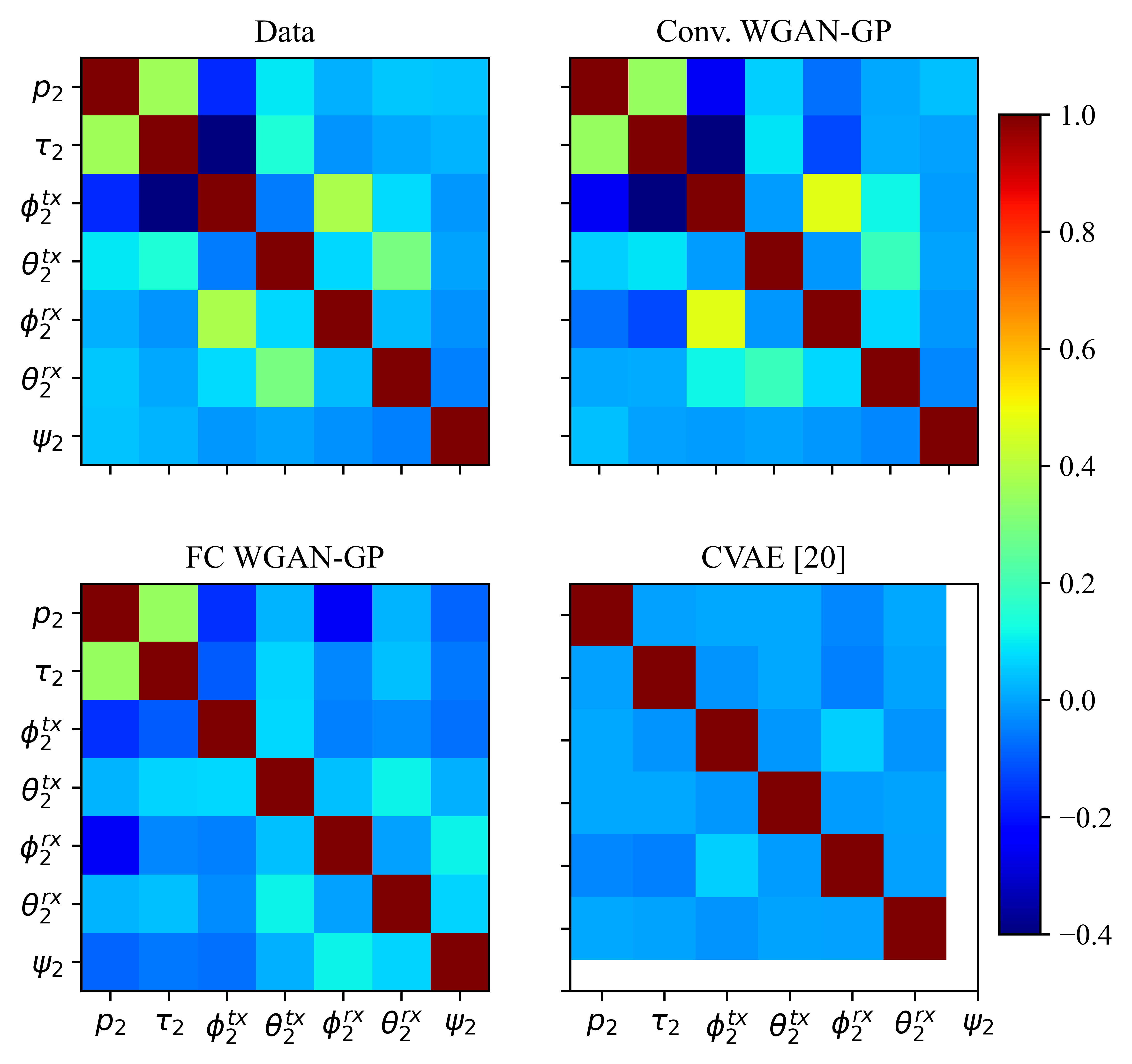} 
\caption{\textcolor{black}{Correlation matrices from data and generative models}}
\label{fig:corr_comp}
\end{figure}
\begin{figure}[!t]
\centering
\includegraphics[width =0.99\columnwidth]{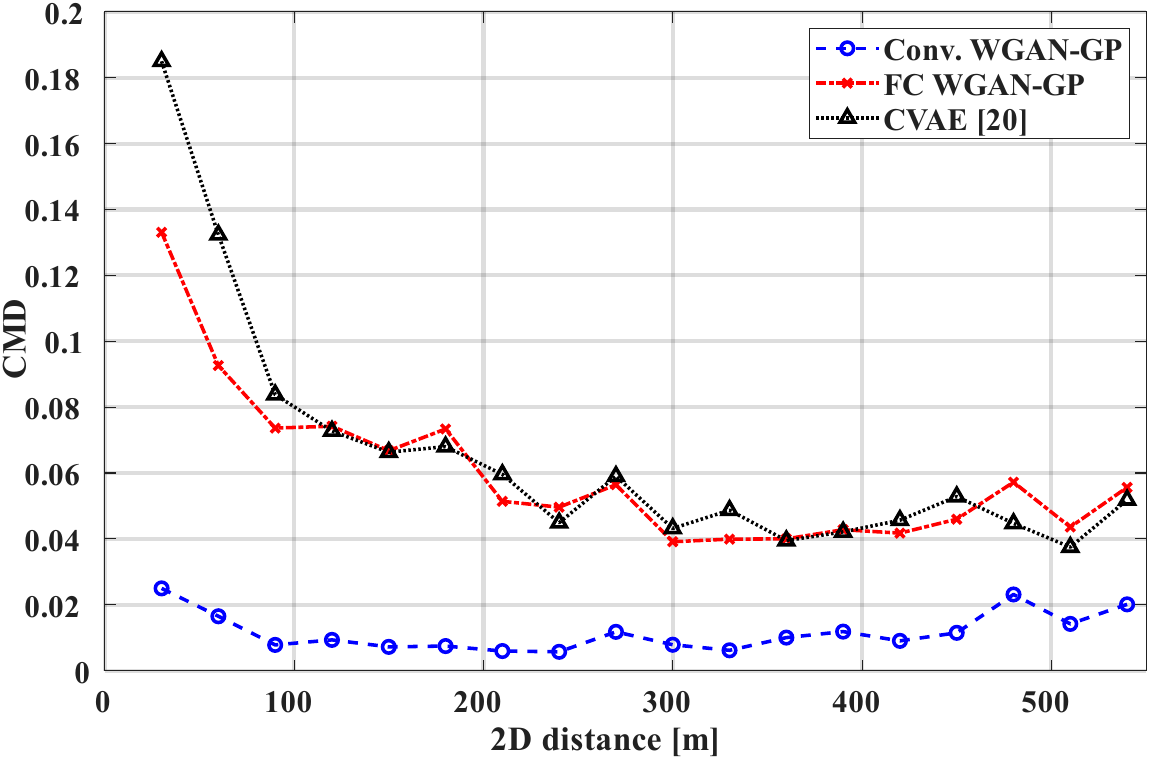} 
\caption{\textcolor{black}{CMDs between the original data and three generative models: Conv. WGAN-GP, FC WGAN-GP, and the CVAE in \cite{xia2022generative}}}
\label{fig:CMD_comp}
\end{figure}
\textcolor{black}{ 
As baseline schemes, we train the WGAN-GP with FC networks and the CVAE~\cite{xia2022generative} with the same data to demonstrate the benefits of employing CNNs in the proposed model.
Fig.~\ref{fig:corr_comp} shows the heatmaps of the correlation matrices $\boldsymbol{R}$ computed from the original data and the outputs of the generative models---Conv.~WGAN-GP, FC WGAN-GP, and CVAE~\cite{xia2022generative}---when the 2D distance and the RX height are fixed at $\SI{150}{m}$ and $\SI{120}{m}$, respectively. 
Note that the CVAE~\cite{xia2022generative} considers only six channel parameters, excluding the phase $\psi_2$, which accounts for the blank entries in its correlation matrix.}

\textcolor{black}{
Fig.~\ref{fig:corr_comp} visually confirms that the correlation matrix generated by the proposed model is more similar to that of the original data than those of the baseline schemes, which implies that employing convolutional neural networks is essential for capturing the correlation between multipath components.
In particular, the FC WGAN-GP and the CVAE~\cite{xia2022generative} largely fail to reproduce the correlation structure of the original data, confirming the limitation of FC-based architectures.
}

In addition, the correlation matrix distance (CMD) between two different correlation matrices can be quantified as described in \cite{herdin2005correlation, giuliani2024spatially}. 
\begin{align*}
    \operatorname{CMD}\left(\boldsymbol{R}_1, \boldsymbol{R}_2\right)=1-\frac{\operatorname{trace}\left(\boldsymbol{R}_1 \boldsymbol{R}_2\right)}{\left\|\boldsymbol{R}_1\right\|_{\mathrm{F}}\left\|\boldsymbol{R}_2\right\|_{\mathrm{F}}} \in[0,1]
\end{align*}
where $\|\cdot\|_F$ denotes Frobenius norm. $\operatorname{CMD}$ becomes close to zero when $\boldsymbol{R}_1$ and $\boldsymbol{R}_2$ are very similar.
\textcolor{black}{
We compute and compare the CMD between the original data and the outputs of each generative model, denoted as $\operatorname{CMD}_{\rm FC}$, $\operatorname{CMD}_{\rm Conv}$, and $\operatorname{CMD}_{\rm CVAE}$ for the FC WGAN-GP, the Conv. WGAN-GP, and the CVAE~\cite{xia2022generative}, respectively.
}

\textcolor{black}{
As shown in Fig.~\ref{fig:CMD_comp}, the FC WGAN-GP and the CVAE~\cite{xia2022generative} result in higher CMD values than the Conv.~WGAN-GP over the entire distance range. In particular, when the TX and the RX are located at close distances, i.e., within $\SI{200}{m}$, the gap between $\operatorname{CMD}_{\rm Conv}$ and the other two metrics becomes most pronounced. Given that a TX and an RX are frequently associated at close distances in practical deployments, this result emphasizes the significance of constructing channel images and employing convolutional layers, which enable the generative model to faithfully reproduce the joint distribution of the original data.}
\textcolor{black}{
\subsection{Interpolation Capability on Unseen Conditions}
The last evaluation step is whether the trained model performs
data synthesis well for conditional inputs that are not included
in the training data. In particular, we want to check how the
trained model performs the statistical interpolation with respect
to the 2D distance $\mathrm{dist}_{\mathrm{2d}}$, since there are
only five conditions of RX heights.
To assess the interpolation capability of the trained model, the dataset is partitioned into training and held-out sets based on the 2D distance conditions, rather than through random splitting.
Specifically, to account for the non-uniform distribution of the 2D distance conditions, the training and held-out sets are constructed from bins defined at the empirical quantiles of the 2D distance conditions, ensuring that each bin contains an equal number of links.

Let $\mathcal{S} = \{\mathrm{dist}_{\mathrm{2d}}^{(1)}, \ldots, \mathrm{dist}_{\mathrm{2d}}^{(M)}\}$, with $|\mathcal{S}| = M$, denote the set of 2D distances of all links.
We place the bin boundaries at the empirical quantiles of
$\mathcal{S}$ such that every bin contains the same number of
links. Let $F$ denote the CDF of $\mathcal{S}$. The bin boundaries are given by
\begin{equation*}
b_i = F^{-1}\!\left(\frac{i}{N}\right), \qquad i = 0, 1, \ldots, N
\end{equation*}
and the $i$-th bin is defined as
\begin{equation*}
\mathcal{B}_i = \left\{\mathrm{dist}_{\mathrm{2d}} \in \mathcal{S}
\,\middle|\, b_{i-1} \le \mathrm{dist}_{\mathrm{2d}} < b_i
\right\}, \quad i = 1, \ldots, N
\end{equation*}
where the upper boundary of the last bin $\mathcal{B}_N$ is
closed so that it includes $\max \mathcal{S}$. By construction, each bin contains $M/N$ links.
Within each bin, the 2D distance conditions are further
partitioned according to a split ratio $p \in (0, 1)$.
For the $i$-th bin, the intra-bin split boundary is
defined as the $p$-quantile of the bin, i.e.,
\begin{equation*}
c_i = F^{-1}\!\left(\frac{i-1+p}{N}\right), \qquad
i = 1, \ldots, N
\end{equation*}
so that the training and held-out sets are given by

\begin{align*}
\mathcal{S}_{\mathrm{train}}(N, p) &= \bigcup_{i=1}^{N}
\left\{\mathrm{dist}_{\mathrm{2d}} \in \mathcal{S} \,\middle|\,
b_{i-1} \le \mathrm{dist}_{\mathrm{2d}} < c_i \right\}, \\
\mathcal{S}_{\mathrm{held-out}}(N, p) &= \bigcup_{i=1}^{N}
\left\{\mathrm{dist}_{\mathrm{2d}} \in \mathcal{S} \,\middle|\,
c_i \le \mathrm{dist}_{\mathrm{2d}} < b_i \right\}.
\end{align*}
This equal-mass partition guarantees that the
training set contains exactly a fraction $p$ of all links,
i.e., $|\mathcal{S}_{\mathrm{train}}| = pM$, independently of
$N$. 
The model is trained only on the data whose 2D distances belong to $\mathcal{S}_{\mathrm{train}}$, and is evaluated on the data whose 2D distances belong to $\mathcal{S}_{\mathrm{held-out}}$.
\begin{figure}[!t]
\centering
\includegraphics[width =0.99\columnwidth]{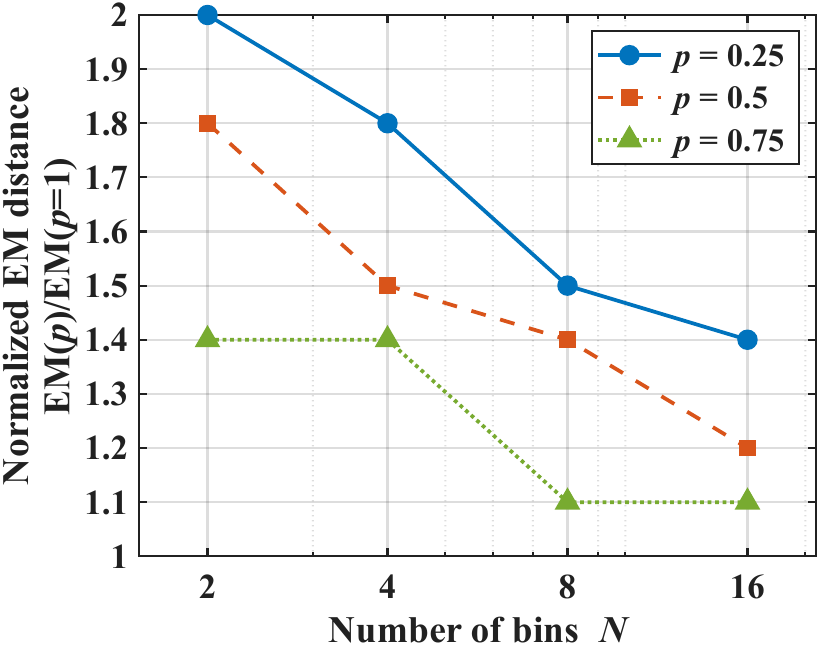} 
\caption{Normalized EM distances vs. number of bins $N$ for different values of $p$}
\label{fig:em_dist_N_p}
\end{figure}


The EM distances are normalized by the EM distance obtained at $p=1$, corresponding to the case in which the model is trained on the entire dataset. Fig.~\ref{fig:em_dist_N_p} shows the normalized EM distance versus the number of bins $N$ for different split ratios $p$. Although the same amount of training data is used across different values of $N$ for a fixed $p$, we observe that the normalized EM distance decreases as $N$ increases, indicating that a finer partitioning of the 2D distance range improves interpolation performance even without additional training data.


Specifically, when the split ratio is large (e.g., $p=0.75$) but the number of bins is small, the normalized EM distance remains relatively high, indicating limited interpolation capability despite the larger amount of training data. In contrast, when the split ratio is small (e.g., $p=0.25$) but the number of bins is large, a comparably low normalized EM distance is achieved. Notably, at $N=16$, the model trained with $p=0.25$ achieves a normalized EM distance of 1.4, matching that obtained with $p=0.75$ at $N=2$, despite using only one-third of the training data. This result suggests that the granularity of the distance partitioning, rather than the amount of training data alone, plays a critical role in achieving good interpolation performance under limited data conditions. 
This finding indicates that the proposed model reliably synthesizes channels under held-out 2D distance conditions, provided that sufficient granularity of the condition split is ensured, thereby supporting its practical value as a data-efficient alternative to directly resampling the ray-tracing database.


}

\section{APPLICATION OF PROPOSED MODEL}
\label{sec:8_application_model}
In this section, we validate the applicability of our model by comparing the simulation results obtained from two different channel models: the deterministic channel model, which was built from ray-tracing data, and our proposed channel model.
\subsection{Simulation Settings}
We consider the coverage of UEs in the dense urban area Herald Square in New York City, when changing the density of BSs. This evaluation is similar to the previous drop-based simulation work carried out in \cite{kang2021millimeter}. We set up the same simulation parameters as in \cite{kang2021millimeter}, and the detailed parameters are listed in Table \ref{tab:table1}.

BSs are deployed according to a homogeneous Poisson point process with a given average intersite distance (ISD), and UEs are dropped at random in the given network area. Each UE is associated with one BS that provides the maximum signal-to-noise ratio (SNR). The \emph{frequency-selective} MIMO channel for uplink (UE $\rightarrow$ BS) $\boldsymbol{H}(f) \in \mathbb{C}^{64 \times 2}$ between BS and UE is built taking the Fourier transform of Eq.~\ref{eq:mimo_chan}. We simply assume optimal beamforming vectors in BSs and UEs: maximum right and left singular vectors of $\boldsymbol{H}(f)$. To eliminate the effect of multipath fading, i.e., frequency-selective fading, for each UE we take the average SNR over available carrier frequencies $f$.
\begin{equation}
    \text{SNR} = \frac{P_{tx}\mathbb{E}_f\left[|\boldsymbol{w}_r^{\text{H}}(f) \boldsymbol{H}(f) \boldsymbol{w}_t(f)|^2\right]}{P_N}
    \label{eq:snr_calc}
\end{equation}
where $P_{tx}$, $\boldsymbol{w}_r \in \mathbb{C}^{64 \times 1}$, and $\boldsymbol{w}_t \in \mathbb{C}^{2 \times 1}$ are beamforming vectors for BSs and UEs, and $P_N$ is noise power. 
\begin{table}[!b]
  \begin{center}
    \caption{Simulation parameters.}
    \label{tab:table1}
    \begin{tabular}{|l|l|} 
    \hline
      \textbf{Parameter} & \textbf{Value}\\
      \hline
      Area (m$^2$)  &  $1120 \times 510 $ \\
      $\ISD$ (m) & $200$, $300$\\
      Height of UEs$~\mathrm{(m)}$ &  $1.6$ \\
      Bandwidth (MHz) & $200$ \\
      Frequency (GHz) & $12$ \\
      UE noise figure (dB) & $7$\\
      UE TX power (dBm) & $23$\\
      BS antenna configuration & $8\times8$ URA\\
      UE antenna configuration & $2\times1$ ULA\\
      Number of BS sectors & 3 \\
      Antenna tilt angle of BSs & $-12^{\circ}$ \\
      Vertical half-power beamwidth ($\theta_{\text{3dB}}$)  & $65^\circ$\\
      Horizontal half-power beamwidth ($\phi_{\text{3dB}}$) & $65^\circ$ \\
      Field pattern of antenna element & 3GPP TR 37.840 \cite{3GPP37840} \\ 
      \hline
    \end{tabular}
  \end{center}
\end{table}
\subsection{Simulation Results}
 With the same simulation settings, the SNR value per UE is calculated for two different channel models: 1) our proposed channel model and 2) the channel based on ray-tracing data.
\begin{figure}[!t]
\centering
\includegraphics[width =0.99\columnwidth]{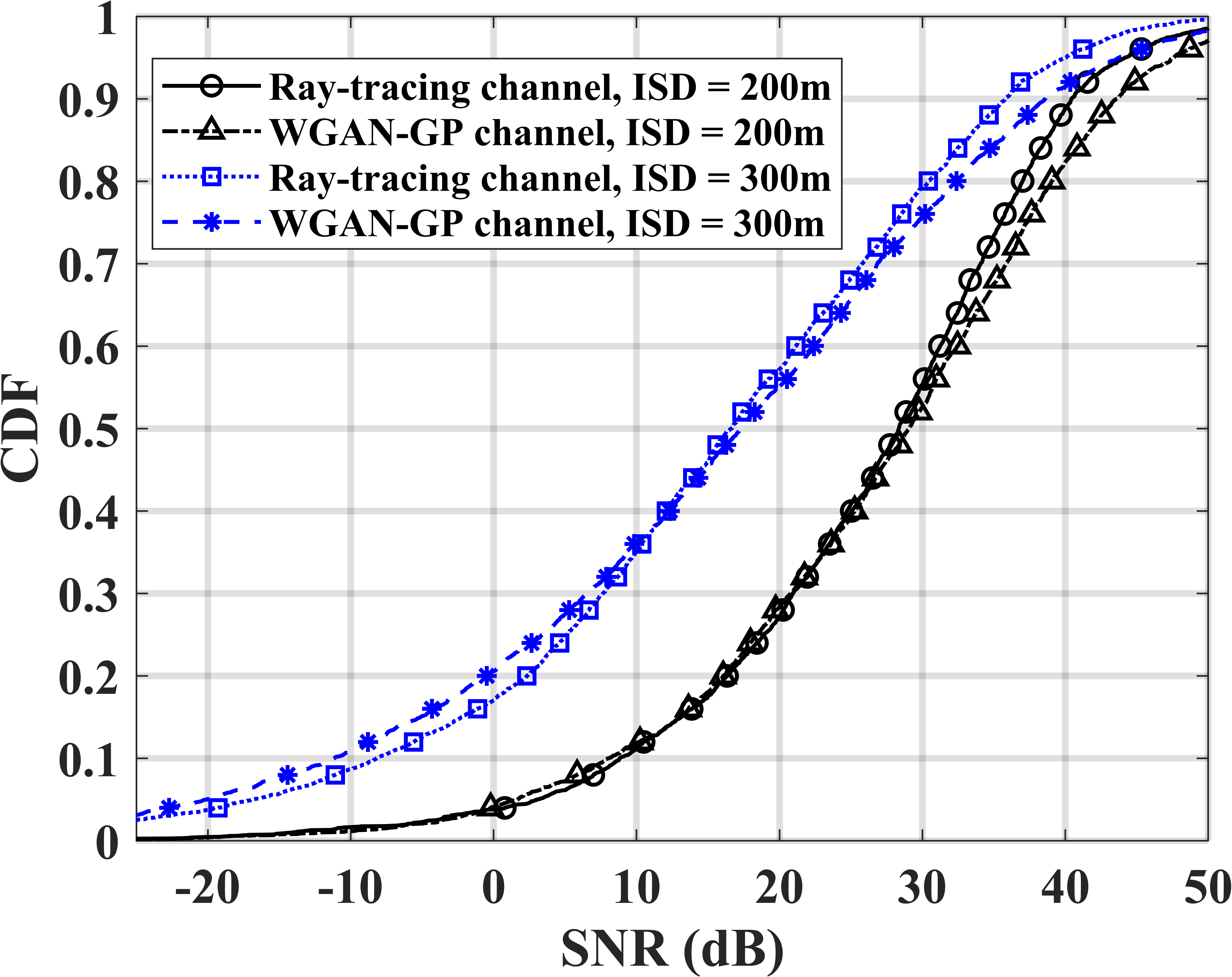} 
\caption{Comparison of SNR distributions with different ISD values}
\label{fig:snr}
\end{figure}
Fig.~\ref{fig:snr} shows simulation results as the SNR CDFs for different channel models when changing ISDs. 
When ISDs are $\SI{200}{m}$ and $\SI{300}{m}$, we observe that the CDFs from the two channel models are almost aligned. The small gaps between CDFs of the same ISDs are due to the lack of randomization of the ray-tracing channel. 

In other words, since the amount of ray-tracing data is limited, considering all possible deployments of BSs is impossible, while our proposed model makes it possible. The simulation results corroborate that our proposed model is sufficient to analyze the system performance, such as coverage and interference, for any given specific area. 
\section{CONCLUSION AND FUTURE WORK}
\label{sec:9_conclusion}
\textcolor{black}{
Developing a theoretical GBSM that accurately captures all statistical characteristics of local scatterers is both highly challenging and computationally expensive. To address this challenge, this work proposes a simplified approach for constructing a GBSM under specific wireless propagation conditions by employing generative neural networks; in particular, the well-known WGAN-GP is adopted. We propose a data-to-image mapping that converts all channel parameters into \emph{channel images}, thereby reducing the complexity of WGAN-GP training and implementation. After training the proposed model, we confirm that the statistics of all channel parameters generated by the WGAN-GP, including the correlations between multipath components, faithfully match the distributions of the original data. In addition, we demonstrate the applicability of the proposed model through a drop-based system-level simulation.}

\textcolor{black}{
Consequently, given wireless channel data corresponding to any 
geometry or a specific site, the proposed data-to-image mapping 
can be combined with other generative models for GBSM construction; 
depending on the required modeling accuracy and the affordable 
implementation complexity, the CVAE or diffusion models can also 
be employed. Moreover, while this work validates the proposed 
framework on a single urban scenario at one carrier frequency, 
the mapping itself is agnostic to the propagation environment. 
Accordingly, a comprehensive comparison across generative model 
families, together with extended validation over held-out receiver heights, multiple carrier frequencies, diverse environmental scenarios, and different ray-tracing configurations, remains an important direction for future work.
}

\bibliographystyle{IEEEtran}
\bibliography{bibl}

\vfill
\vspace{11pt}


\end{document}